\documentclass[10pt,aps,showpacs,nofootinbib,prd,aps,floats,
               amsmath,amssymb,amsfonts,floatfix,twocolumn]{revtex4-1}
\usepackage{amsmath, amssymb}
\usepackage{multirow}
\usepackage{paralist}
\usepackage{slashed}
\usepackage{hyperref} 
\newcommand{\mathsym}[1]{{}}

\usepackage{graphicx}
\usepackage{amsmath}
\usepackage{amssymb}
\usepackage{mathrsfs}
\newsavebox{\PSLASH}
 \sbox{\PSLASH}{$p$\hspace{-1.8mm}/}
 
\renewcommand{\theequation}{\thesection.\arabic{equation}}
\newcounter{saveeqn}
\newcommand{\add}{\addtocounter{equation}{1}}
\newcommand{\alphaeqn}{\setcounter{saveeqn}{\value{equation}}%
\setcounter{equation}{0}%
\renewcommand{\theequation}{\mbox{\thesection.\arabic{saveeqn}{\alpha{equation}}}}}
\newcommand{\reseteqn}{\setcounter{equation}{\value{saveeqn}}%
\renewcommand{\theequation}{\thesection.\arabic{equation}}}

 \newsavebox{\notrightarrow}
 \sbox{\notrightarrow}{$\to$\hspace{-4mm}/}
 
 \newsavebox{\PARTIALSLASH}
 \sbox{\PARTIALSLASH}{$\partial$\hspace{-1.6mm}/}
 
 \newsavebox{\ASLASH}
 \sbox{\ASLASH}{$A$\hspace{-2.1mm}/}
 
 \newsavebox{\KSLASH}
 \sbox{\KSLASH}{$k$\hspace{-1.8mm}/}
 
 \newsavebox{\LSLASH}
 \sbox{\LSLASH}{$\ell$\hspace{-1.8mm}/}
 
 \newsavebox{\QSLASH}
 \sbox{\QSLASH}{$q$\hspace{-1.8mm}/}
 
 \newsavebox{\DSLASH}
 \sbox{\DSLASH}{$D$\hspace{-2.2mm}/}
 
 \newsavebox{\DbfSLASH}
 \sbox{\DbfSLASH}{${\mathbf D}$\hspace{-2.8mm}/}
 
 \newsavebox{\DELVECRIGHT}
 \sbox{\DELVECRIGHT}{$\stackrel{\rightarrow}{\partial}$}
 
 \newcommand{\blue}{\IfColor{\textCadetBlue}{}}
\newcommand{\black}{\IfColor{\textBlack}{}}
\newcommand{\red}{\IfColor{\textRed}{}}
\newcommand{\green}{\IfColor{\textOliveGreen}{}}
\newcommand{\lil}{\IfColor{\textRedViolet}{}}

\newcommand{\bs}{\boldsymbol}

\usepackage{orcidlink}

\begin{document}

\title{Relativistic magnetohydrodynamics of a spinful and vortical fluid: Entropy current analysis}
\author{M. Kiamari $^{a}$}\email{mkiamari@ipm.ir}
\author{N. Sadooghi $^{b}$}\email{Corresponding author: sadooghi@physics.sharif.ir}
\author{M. Sedighi Jafari $^{b}$}\email{mahdi.sedighi@physics.sharif.edu}
\affiliation{$^{a}$School of Particles and Accelators, IPM, P.O. Box 19395-5531, Tehran, Iran}
\affiliation{$^{b}$Department of Physics, Sharif University of Technology,
P.O. Box 11155-9161, Tehran, Iran}
\begin{abstract}
We generalize a recently introduced formulation of relativistic spinful and vortical fluid to relativistic magnetohydrodynamics (MHD). We refer to it as the "Spinful-Vortical MHD" (SVMHD). The aim is to scrutinize the interplay between the vorticity, magnetic field, and spin, which is treated as a quantum object, in contrast to other formulations of spin hydrodynamics. To this purpose, we first  perform a standard entropy current analysis up to first-order gradient expansion, $\mathcal{O}\left(\partial\right)$ as well as $\mathcal{O}\left(\hbar\partial\right)$, where $\hbar$ is the Planck constant. In contrast to alternative formulations of spin MHD, in the absence of vorticity, the zeroth-order energy-momentum tensor includes an additional magneto-vorticity mixed term and reduces, as expected, to the energy-momentum tensor of MHD. We show that in the first-order of gradient expansion, $36$ dissipative transport coefficients appear. They satisfy certain constraints that guarantee the positive definiteness of the entropy production rate. We then modify the formulation of SVMHD by replacing the magnetic part of the thermal vorticity tensor with its electric part. Carrying out the same analysis as in the standard formulation, we show that in this case, the first-order constitutive relations consist of $11$ nondissipative Hall-like coefficients, apart from $25$ dissipative coefficients. This difference arises from different behavior of the electric and magnetic part of the thermal vorticity under time-reversal transformation. 
\end{abstract}
\maketitle
\section{Introduction}\label{sec1}
\setcounter{equation}{0}
Relativistic hydrodynamics (RH) \cite{rezzolla-book, rischke-book} is accepted as a paradigm for the description of the evolution of the Quark-Gluon Plasma (QGP) created in relativistic heavy-ion collisions (HICs) at the Relativistic Heavy Ion Collider (RHIC) and Large Hadron Collider (LHC) \cite{heinz2013, heller2017, shen2020}. The QGP is characterized by extreme temperature, high energy density, and, in particular, large vorticity. In \cite{becattini2016}, the vorticity of QGP is estimated to be $0.02$ fm$^{-1}\approx 10^{22}$ Hz. The creation of this extreme vorticity in the QGP produced in HIC is confirmed in the measurement of global spin polarization of $\Lambda$ hyperons at RHIC \cite{star2017}. Moreover,
the measurements of hyperon polarization and vector meson spin alignment in HICs at RHIC and LHC \cite{star2018, star2019,star2020, star2022,alice2022}, indicate a strong spin polarization of quarks and gluons in the QGP and confirm earlier theoretical works \cite{theospin2004-1, theospin2004-2,betz2007,becattini2007,huang2011,becattini2013} (see also recent reviews \cite{rev-wang2020, rev-huang2020, rev-lisa2020, becattini2022} and the references therein). The idea is that two colliding nuclei at finite impact parameters with a large relative angular momentum give rise to a finite vorticity in the QGP. The resulting vorticity induces a spin polarization in the direction of the local vorticity vector \cite{becattini2022}. This is in analogy to the Barnett effect, according to which spinning a ferromagnet changes its magnetization \cite{rev-lisa2020}. The key point here is the transfer of angular momentum from orbital to spin degrees of freedom \cite{becattini2021}.
\par
The above-mentioned experimental efforts call for a consistent generalization of RH to a RH for a spinful and vortical fluid that enables a rigorous study of the coupling of spin and vorticity.
Numerous theoretical attempts have been made in this direction. The resulting RH is known as relativistic spin hydrodynamics\cite{florkowski2017,becattini2018,florkowski2018,hattori2019,fukushima2019,florkowski2020,
gale2020,stephanov2020,nora2022,mishra2022, yarom2021,wang2021,ryblewski2022,hu2022,pu2023,nora2023,becattini2023}.
Recently in \cite{huang2022}, a formalism is developed where vorticity is counted as zeroth order in derivative, although it contains a first-order gradient of fluid velocity and is thus comparable with the dissipative part of the energy-momentum tensor. Another characteristic of this formalism is that the spin density is at the leading order in derivative, but is suppressed by the Planck constant $\hbar$. The latter is considered because of the quantum nature of spin. This contrasts with other formulations of relativistic spin hydrodynamics, where spin and vorticity are considered to be of the same order as the viscous corrections to hydrodynamics.
\par
Assuming generic forms for a rank two energy-momentum tensor $\Theta^{\mu\nu}$, a rank three and totally antisymmetric spin tensor $\Sigma^{\mu\nu\rho}$, and a vector current $J^{\mu}$ and performing an appropriate entropy current analysis up to the first-order derivative expansion, the authors of \cite{huang2022} determine $\Theta^{\mu\nu}$ of an ideal spinful fluid in terms of the energy density $\epsilon$, the
four-vector of the thermal vorticity $\omega^{\mu}$,\footnote{The four-vector $\omega^{\mu}$ is the dual vector to the thermal vorticity tensor $\omega_{\mu\nu}\equiv \partial_{[\mu}\beta_{\nu]}$, with $\beta^{\mu}\equiv \beta u^{\mu}$. Here, $\beta$ is the inverse temperature and $u^{\mu}$ the velocity of the fluid (see \ref{sec2A} for more details).} parallel as well as perpendicular pressures $p_{\|}$ and $p_{\perp}$. The subscripts $\|$ and $\perp$ indicate parallel and perpendicular directions with respect to the direction of vorticity. Whereas the $p_{\|}$ does not receive any correction in zeroth order gradient expansion, $p_{\perp}$ consists of a term proportional to parallel spin density $\sigma_{\|}$, with a parallel spin potential $\mu_{\|}$ as the proportionality factor. Here,  $\sigma_{\|}$ and $\mu_{\|}$ arise from a certain decomposition of two vectors $\sigma^{\mu}$ and $\mu^{\mu}$ into parallel and perpendicular parts with respect to $\omega^{\mu}$.\footnote{Four-vectors $\sigma^{\mu}$ and $\mu^{\mu}$  are dual vectors corresponding to spin and spin potential tensors $\sigma^{\mu\nu}$ and $\mu^{\mu\nu}$, often used in the literature.} Whereas the parallel parts of $\sigma^{\mu}$ and $\mu^{\mu}$ are assumed to be of order $\mathcal{O}(\hbar)$ and $\mathcal{O}(1)$, the perpendicular parts are of order $\mathcal{O}(\hbar\partial)$ and $\mathcal{O}(\partial)$, respectively. The latter is thus negligible in the ideal case. According to this result, the vorticity of the fluid induces an anisotropy in the pressure.
Apart from the energy-momentum tensor of the ideal spinful and vortical fluid, the constitutive relations of this model are determined by making use of the second law of thermodynamics. It is shown that apart from seventeen anisotropic dissipative transport coefficients, seven Hall-like transport coefficients appear, which do not contribute to entropy production. Furthermore, the inequalities satisfied by the seventeen dissipative coefficients are derived.
\par
In the present paper, we generalize the formulation introduced in \cite{huang2022} to relativistic magnetohydrodynamics (MHD). We refer to this novel formulation as spinful vortical MHD (SVMHD).
The motivation behind the extension of \cite{huang2022} to MHD is the production of strong magnetic fields up to $10^{18}$-$10^{20}$ G \cite{warringa2007,skokov2009} in noncentral HICs at RHIC and LHC, that apart from vorticity of the QGP contribute to the polarization of the medium.
The effect of external magnetic fields on various properties of the quark matter is intensively studied in the literature (see e.g. \cite{rev-huang2015,sadooghi2010,  sadooghi2012,sadooghi2013,sadooghi2014,sadooghi2015,taghinavaz2016,sadooghi2020} and references therein). In \cite{sadooghi2019}, e.g., the fact that a uniform magnetic field creates an anisotropic pressure in the direction perpendicular to the magnetic field is used to develop anisotropic relativistic hydrodynamics. This is then used to study the time evolution of thermodynamic variables in and out of equilibrium. In \cite{rischke2009, rischke2011,sadooghi2016}, the effect of magnetization and electric polarization on the dissipative and anomalous transport coefficients of a chiral fluid is studied. It is shown, that the presence of external magnetic fields leads to five shear viscosities and two bulk viscosities, including also Hall-like transport coefficients. New developments in relativistic dissipative MHD are presented in \cite{iqbal2016, hongo2020, hongo2022}.
\par
A combination of spin hydrodynamics with MHD is thus of great importance. In \cite{florkowski2022}, a formulation of spin MHD is presented in the framework of the kinetic theory of massive spin $1/2$ particles. It is, in particular, shown that the coupling between spin polarization and magnetic field occurs at gradient order. Related works in the same direction explore the time evolution of the spin polarization tensor by solving the spin MHD equations  \cite{shokri2022,buzzegoli2022,pu2022}.
\par
The organization of the paper is as follows: In Sec. \ref{sec2A},
we first introduce the SVMHD model by generalizing the method introduced in \cite{huang2022} to the case of a spinful and vortical relativistic fluid in the presence of electromagnetic fields and derive  the four-divergence of the corresponding entropy current. In Sec. \ref{sec2}, we then perform an appropriate entropy current analysis up to first-order in gradient expansion. In \ref{sec2B}, we derive the energy-momentum tensor for the case when the vorticity and magnetic field are perpendicular to each other. We show, in particular, that a mixed term including the unit vectors in the direction of the vorticity and magnetic field appears in the energy-momentum tensor of an ideal spinful and vortical fluid. This mixed term appears also in the case when the vorticity and magnetic field are parallel (see Appendix \ref{appA}). This result is in contrast to the results presented in \cite{florkowski2022}, where such a mixed term appears in $\mathcal{O}(\partial)$. In Sec. \ref{sec2C}, we derive the transport coefficients in the first-order gradient expansion.
We show that in this model $36$ dissipative coefficients arise. They consists of four thermal conductivities, five rotational, ten bulk, four shear, eight cross viscosities, and five electric resistivities. All these coefficients contribute to entropy production and satisfy certain inequalities that guarantee the second law of thermodynamics (see Sec. \ref{sec2C}). Up to this stage, we use the magnetic part of the thermal vorticity tensor $\omega^{\mu\nu}$, i.e. $\omega^{\mu\nu}=\epsilon^{\mu\nu\rho\sigma}\omega_{\rho}u_{\sigma}$. In Sec. \ref{sec2D}, we replace $\omega^{\mu}$ with the electric part of the thermal vorticity (spin polarization tensor) $\kappa^{\mu}$, arising from the decomposition of the second-rank spin polarization tensor $\omega^{\mu\nu}$ into its electric and magnetic part,  $\omega^{\mu\nu}=\kappa^{\mu}u^{\nu}-\kappa^{\nu}u^{\mu}$. We derive the energy-momentum tensor for this new formulation up to first-order in gradient expansion. It turns out that the energy-momentum tensor in the ideal case is given by replacing $\omega^{\mu}$ with $\kappa^{\mu}$. In the first-order derivative expansion, in contrast to the previous case discussed in \ref{sec2C}, eleven Hall-like coefficients appear that do not contribute to entropy production. The reason for this effect is the different behavior of $\omega^{\mu}$ and $\kappa^{\mu}$, the magnetic and electric part of $\omega^{\mu\nu}$, with respect to time reversal, which leads to different Onsager conditions for $\omega^{\mu}$ and $\kappa^{\mu}$. In Sec. \ref{sec4}, we present the summary of our results and a brief outlook into future works.\\
\par\noindent\textbf{\textit{Conventions and notations:}}
In this paper, we use the mostly negative metrics defined by $g_{\mu\nu}\equiv \mbox{diag}\left(1,-1,-1,-1\right)$. The totally antisymmetric tensor $\epsilon^{\mu\nu\rho\sigma}$ satisfies $\epsilon^{0123}=-\epsilon_{0123}=+1$. For  $X\in \{b,\omega, \sigma,\mu\}$, the rank-two tensor $X^{\mu\nu}$ is defined by  $X^{\mu\nu}\equiv \epsilon^{\mu\nu\rho\sigma}X_{\rho}u_{\sigma}$. This leads to $u_{\mu}X^{\mu\nu}=0$.
Using the properties of $\epsilon^{\mu\nu\rho\sigma}$ and the definition of $X^{\mu\nu}$, we have $X^{\mu}=-\frac{1}{2}\epsilon^{\mu\nu\rho\sigma}u_{\nu}X_{\rho\sigma}$ for $X\in \{b,\omega, \sigma,\mu\}$. This leads immediately to $u_{\mu}X^{\mu}=0$.
The symmetric part of the tensor $X^{\mu\nu}$ is defined by $X^{(\mu\nu)}\equiv \frac{1}{2}\left(X^{\mu\nu}+X^{\nu\mu}\right)$ and its antisymmetric part by
$X^{[\mu\nu]}\equiv \frac{1}{2}\left(X^{\mu\nu}-X^{\nu\mu}\right)$.
The dual $\tilde{X}^{\mu\nu}$ is defined by
$\tilde{X}^{\mu\nu}\equiv \frac{1}{2}\epsilon^{\mu\nu\rho\sigma}X_{\rho\sigma}$.
We use the notation $a\cdot b\equiv a_{\mu} b^{\mu}$.
 The unit vectors $u_{\mu}, b_{\mu},$ and $\omega_{\mu}$ satisfy
$u\cdot u=+1$, $b\cdot b=-1$, and $\omega\cdot\omega=-1$. Moreover, we have $u\cdot b=u\cdot \omega=0$.
We finally define $\Delta^{\mu\nu}\equiv g^{\mu\nu}-u^{\mu}u^{\nu}$ and $\Xi^{\mu\nu}\equiv \Delta^{\mu\nu}+b^{\mu}b^{\nu}+\omega^{\mu}\omega^{\nu}$.
\section{The SVMHD Model}\label{sec2A}
\setcounter{equation}{0}
To formulate the magnetohydrodynamics of a charged, rotating, and spinful fluid, we start with the homogeneous Maxwell equation
\begin{eqnarray}\label{N1a}
\partial_{\mu}\tilde{F}^{\mu\nu}=0,
\end{eqnarray}
and the conservation laws corresponding to the electromagnetic current, energy-momentum, and total angular momentum,
\begin{eqnarray}\label{N2a}
\partial_{\mu}J^{\mu}=0,\quad
\partial_{\mu}\Theta^{\mu\nu}=0, \quad\partial_{\mu}J^{\mu\nu\rho}=0.
\end{eqnarray}
Here, $\tilde{F}_{\mu\nu}$ is the dual (electromagnetic) field strength tensor and $J^{\mu}$ is the electromagnetic current. Moreover, $\Theta^{\mu\nu}$ and $J^{\mu\nu\rho}=L^{\mu\nu\rho}+\Sigma^{\mu\nu\rho}$ are the energy-momentum and the total angular momentum tensors. The latter consists of an orbital  $L^{\mu\nu\rho}$ and a spin part $\Sigma^{\mu\nu\rho}$. It is known that $L^{\mu\nu\rho}=x^{\nu}\Theta^{\mu\rho}-x^{\rho}\Theta^{\mu\nu}$. Having this in mind, the conservation laws \eqref{N2a} lead to
\begin{eqnarray}\label{N3a}
\partial_{\mu}\Sigma^{\mu\nu\rho}=-2\Theta^{[\nu\rho]},
\end{eqnarray}
where $\Theta^{[\nu\rho]}$ is the antisymmetric part of the energy-momentum tensor. Apart from the above conservations laws, we need to define a number of thermodynamical variables. In the Landau frame, the number density $n$, the energy density, consisting of the energy density of the fluid $\epsilon$ and the energy density of an electromagnetic field $B^{2}/2$, and the spin density tensor $\sigma^{\mu\nu}$ are defined by
\begin{eqnarray}\label{N4a}
u_{\mu}J^{\mu}&=&n,\nonumber\\
u_{\nu}\Theta^{(\mu\nu)}&=&\left(\epsilon+B^2/2\right)u^{\mu},\nonumber\\
u_{\mu}\Sigma^{\mu}_{~\nu\rho}&=&-\sigma_{\nu\rho}.
\end{eqnarray}
Here, $u^{\mu}$ is the fluid velocity, which is defined by $u^{\mu}=\gamma(1,\bs{v})$ with $\gamma=\left(1-\bs{v}^{2}\right)^{-1/2}$. It satisfies $u_{\mu}u^{\mu}=1$.
Moreover, $\Theta^{(\mu\nu)}$ is the symmetric part of the energy-momentum tensor.
Using the first law of thermodynamics for a system consisting of charged particles with spin and in the presence of a constant magnetic field,
\begin{eqnarray}\label{N5a}
Tds=d\epsilon-\frac{1}{2}\mu^{\mu\nu}d\sigma_{\mu\nu}-\mu dn+MdB,
\end{eqnarray}
we arrive at appropriate definitions of the inverse temperature $\beta=T^{-1}$, the spin potential $\mu^{\mu\nu}$, the chemical potential $\mu$, and the magnetization $M$,
\begin{eqnarray}\label{N6a}
\beta&=&\left(\frac{\partial s}{\partial\epsilon}\right)_{\sigma^{\mu\nu},n, B}, \qquad
\beta\mu^{\mu\nu}=-2\left(\frac{\partial s}{\partial\sigma_{\mu\nu}}\right)_{\epsilon,n, B},\nonumber\\
\beta\mu&=&-\left(\frac{\partial s}{\partial n}\right)_{\sigma^{\mu\nu},\epsilon, B},\qquad
\beta M=\left(\frac{\partial s}{\partial B}\right)_{\sigma^{\mu\nu},\epsilon, n}.\nonumber\\
\end{eqnarray}
To start the entropy current analysis and to arrive at the constitutive relations of a rotating spinful fluid in the presence of electromagnetic fields, we use the following decomposition of $J^{\mu}, \Theta^{\mu\nu}, \tilde{F}^{\mu\nu}$, and $\Sigma^{\mu\nu\rho}$ into their ideal and dissipative parts \cite{huang2022}
\begin{eqnarray}\label{N7a}
J^{\mu}&=&nu^{\mu}+\delta J^{\mu},\nonumber\\
\Theta^{\mu\nu}&=&\left(\epsilon+p_{1}+\alpha_1 B^{2}\right)u^{\mu}u^{\nu}-g^{\mu\nu}\left(p_{1}+\alpha_2 B^{2}\right)\nonumber\\
&&+p_{2}{\omega}^{\mu}{\omega}^{\nu}+p_{3}\omega^{\mu\nu}+p_{4}b^{\mu}b^{\nu}+p_{5}b^{\mu\nu}\nonumber\\
&&+2p_{6}b^{\left(\mu\right.}{\omega}^{\left.\nu\right)}
+u^{\mu}\delta q^{\nu}-u^{\nu}\delta q^{\mu}+\delta\Theta^{\mu\nu},\nonumber\\
\tilde{F}^{\mu\nu}&=&B^{\mu}u^{\nu}-B^{\nu}u^{\mu}+\delta\tilde{F}^{\mu\nu},\nonumber\\
\Sigma^{\mu\nu\rho}&=&\epsilon^{\mu\nu\rho\lambda}\left(\sigma_{\lambda}+u_{\lambda}\delta\sigma\right).
\end{eqnarray}
Here, the four-vector of the thermal vorticity $\omega^{\mu}$ is the dual vector of the thermal vorticity tensor $\omega^{\mu\nu}\equiv \partial_{[\mu}\beta_{\nu]}$, with $\beta^{\mu}\equiv u^{\mu}/T$. It satisfies $\omega_{\mu}\omega^{\mu}=-1$\footnote{In this paper, we used the notation $\omega^{\mu}$ for the normalized thermal vorticity vector. This is in contrast to the notation used in \cite{huang2022}, where the same vector is denoted by $\hat{\omega}^{\mu}$.} as well as $u^{\mu}\omega_{\mu}=0$. The Minkowski metric $g^{\mu\nu}$ reads $g^{\mu\nu}=\mbox{diag}\left(1,-1,-1,-1\right)$ and $B^{\mu}$, the four-vector of the magnetic field, is defined in terms of the field strength tensor $F^{\mu\nu}$ as $B^{\mu}=\frac{1}{2}\epsilon^{\mu\nu\rho\sigma}F_{\nu\rho}u_{\sigma}$. In the rest frame of the fluid, $B^{\mu}=\left(0,\boldsymbol{B}\right)$. The unit vector in the direction of the magnetic field is given by $b^{\mu}\equiv \frac{B^{\mu}}{B}$ and satisfies $b_{\mu}b^{\mu}=-1$ as well as $u^{\mu}b_{\mu}=0$. Similarly, it is possible to introduce the magnetic polarization vector $M^{\mu}$, that in the rest frame of the fluid reads $M^{\mu}=\left(0,\bs{M}\right)$, with $\bs{M}=\chi_{m}\bs{B}$ \cite{rischke2009,sadooghi2016}. Here, $\chi_{m}$ is the magnetic susceptibility.
\par
Let us notice that, in the Landau frame, the dissipative parts $\delta J^{\mu}, \delta q^{\mu}, \delta\Theta^{\mu\nu}, \delta\tilde{F}^{\mu\nu}$, and $\delta \sigma$, they satisfy $u_{\mu}\delta J^{\mu}=0, u_{\mu}\delta q^{\mu}=0, u_{\mu}\delta\Theta^{\mu\nu}=0$, and $u_{\mu}\delta\tilde{F}^{\mu\nu}=0$.
Using these properties and
\begin{eqnarray}\label{N8a}
u^{\rho}\partial_{\mu}\Sigma^{\mu}_{~\nu\rho}=-2u^{\rho}\Theta_{[\nu\rho]},
\end{eqnarray}
arising from \eqref{N3a}, $\delta q^{\mu}$ can be given in terms of the spin vector $\sigma^{\mu}$ and the dissipative part of $\Sigma^{\mu\nu\rho}$, $\delta\sigma$, as
\begin{eqnarray}\label{N9a}
\delta q^{\mu}=-\frac{1}{2}\epsilon^{\mu\nu\rho\lambda}u_{\rho}\partial_{\nu}\left(\sigma_{\lambda}+u_{\lambda}\delta\sigma\right).
\end{eqnarray}
We will determine $\delta \sigma$ later. In the ansatz which is introduced for $\Theta^{\mu\nu}$, $p_{i}, i=1,2,\cdots 6$ are unknown quantities. In what follows, they will be determined by making use of standard symmetry argumentation. The parameters $\alpha_{1}$ and $\alpha_{2}$ are fixed by multiplying the symmetric part of $\Theta^{\mu\nu}$ with $u_{\mu}$ and requiring that the Euler relation of thermodynamics $\epsilon+p=Ts+\mu n$ is independent of $B$ [see \eqref{N19a}]. They are thus given by $\alpha_{1}=1$ and $\alpha_{2}=1/2$.
\par
To determine $ds$ from \eqref{N5a}, we insert $J^{\mu}, \Theta^{\mu\nu}, $ and $\Sigma^{\mu\nu\rho}$ into the conservation laws \eqref{N2a}. Contracting then the resulting expressions with $\beta_{\nu}, -\frac{1}{2}\mu^{\nu\rho}, \mu\beta$, and $\beta M$, we arrive after some work at
\begin{widetext}
\begin{eqnarray}\label{N10a}
\beta D\epsilon&=&-\beta\theta\left(\epsilon+p_{1}\right)+\left(p_{2} {\omega}^{\mu} {\omega}^{\nu}+p_{3}\omega^{\mu\nu}+\left(p_{4}+B^2\right)b^{\mu}b^{\nu}+p_{5}b^{\mu\nu}+2p_{6}b^{\left(\mu\right.} {\omega}^{\left.\nu\right)}\right)\partial_{\mu}\beta_{\nu}+\delta q^{\nu}\left(D\beta_{\nu}-\partial_{\nu}\beta\right)\nonumber\\
&&+\partial_{\mu}\left(\beta \delta q^{\mu}\right)+\delta\Theta^{\mu\nu}\partial_{\mu}\beta_{\nu},\nonumber\\
-\frac{1}{2}\beta\mu^{\nu\rho}D\sigma_{\nu\rho}&=&-\mu^{\mu}\sigma_{\lambda}\partial_{\mu}\beta^{\lambda}+\partial_{\mu}\left(\beta\mu^{\mu}\delta\sigma\right)+\frac{1}{2}\epsilon^{\mu\nu\rho\lambda}\delta\sigma u_{\lambda}\partial_{\mu}\left(\beta\mu_{\nu\rho}\right)-p_{3}\beta\mu^{\nu\rho}\epsilon_{\nu\rho}-\beta\mu^{\nu\rho}\delta\Theta_{[\nu\rho]}-p_{5}\beta \mu_{\nu\rho}b^{\nu\rho}\nonumber\\
-\mu\beta Dn&=&\beta\mu n\theta+\partial_{\mu}\left(\mu\beta\delta J^{\mu}\right)-\delta J^{\mu}\partial_{\mu}\left(\beta \mu\right),\nonumber\\
\beta MDB&=&-\beta MB\theta-MB b^{\mu}b^{\nu}\partial_{\mu}\beta_{\nu}+\delta\tilde{F}^{\mu\nu}\partial_{\mu}\left(\beta M_{\nu}\right)-\partial_{\mu}\left(\beta M_{\nu}\delta\tilde{F}_{\mu\nu}\right),
\end{eqnarray}
\end{widetext}
where $D\equiv u^{\mu}\partial_{\mu}$ and $\theta\equiv \partial_{\mu}u^{\mu}$.
Using $s^{\mu}=su^{\mu}+\delta s^{\mu}$ and plugging \eqref{N10a} into
\begin{eqnarray}\label{N11a}
Ds=\beta D\epsilon-\frac{1}{2}\beta\mu^{\nu\rho}D\sigma_{\nu\rho}-\mu\beta Dn+\beta MDB, \nonumber\\
\end{eqnarray}
arising from \eqref{N5a}, we obtain
\begin{widetext}
\begin{eqnarray}\label{N12a}
\partial_{\mu}s^{\mu}&=&\theta\big[s-\beta\left(\epsilon+p_{1}-\mu n+MB\right)\big]+\left(p_{2}  {\omega}^{\mu}  {\omega}^{\nu}-\mu^{\mu}\sigma^{\nu}+\left(p_{4}+B^2-MB\right)b^{\mu}b^{\nu}\right.\nonumber\\
&&\left.+p_{5}b^{\mu\nu}+2p_{6}b^{\left(\mu\right.}  {\omega}^{\left.\nu\right)}\right)\partial_{\mu}\beta_{\nu}+p_{3}\omega^{\mu\nu}\left(\partial_{\mu}\beta_{\nu}-\beta\mu_{\mu\nu}\right)+\delta\Theta^{(\mu\nu)}\partial_{(\mu}\beta_{\nu)}+\delta\Theta^{[\mu\nu]}\left(\partial_{[\mu}\beta_{\nu]}-\beta\mu_{\mu\nu}\right)\nonumber\\
&&+\partial_{\mu}\left(\delta s^{\mu}+\beta\delta q^{\mu}+\beta\mu^{\mu}\delta\sigma+\mu\beta\delta J^{\mu}-\beta M_{\nu}\delta\tilde{F}^{\mu\nu}\right)-\delta J^{\mu}\partial_{\mu}\left(\beta\mu\right)-p_{5}\mu_{\nu\rho}b^{\nu\rho}-\delta\sigma\partial_{\mu\perp}\left(\bar{\omega}^{\mu}+\beta\mu^{\mu}\right)\nonumber\\
&&+\frac{1}{2}\epsilon^{\mu\nu\rho\lambda}u_{\rho}\partial_{\mu}\sigma_{\lambda}\left(D\beta_{\nu}-\partial_{\nu}\beta\right)+\delta\tilde{F}^{\mu\nu}\partial_{\mu}(\beta M_{\nu}),
\end{eqnarray}
\end{widetext}
where $\partial_{\mu\perp}\equiv \left(\delta_{\mu}^{~\nu}-u_{\mu}u^{\nu}\right)\partial_{\nu}$. We also have introduced the vorticity $\tilde{\omega}^{\mu}\equiv \frac{1}{2}\epsilon^{\mu\nu\alpha\beta}u_{\nu}\partial_{\alpha}u_{\beta}$ and $\bar{\omega}^{\mu}\equiv \beta \tilde{\omega}^{\mu}$ and used the relation $\tilde{\omega}^{\rho}Du_{\rho}=-\frac{1}{2}\partial_{\mu}\tilde{\omega}^{\mu}$ \cite{sadooghi2016} to arrive at
\begin{eqnarray}\label{N13a}
\bar{\omega}^{\mu}Du_{\mu}=-\partial_{\mu\perp}\bar{\omega}^{\mu}+T\bar{\omega}^{\mu}\partial_{\mu}\beta.
\end{eqnarray}
Using further $\delta q^{\nu}$ from \eqref{N9a}, we obtain
\begin{eqnarray}\label{N14a}
\lefteqn{\delta q^{\nu}\left(D\beta_{\nu}-\partial_{\nu}\beta\right)+\frac{1}{2}\epsilon^{\mu\nu\rho\lambda}\delta\sigma u_{\lambda}\partial_{\mu}\left(\beta \mu_{\nu\rho}\right)}\nonumber\\
&=&\frac{1}{2}\epsilon^{\mu\nu\rho\lambda}u_{\rho}\partial_{\mu}\sigma_{\lambda}\left(D\beta_{\nu}-\partial_{\nu}\beta\right)-\delta\sigma\partial_{\mu\perp}\left(\bar{\omega}^{\mu}+\beta\mu^{\mu}\right). \nonumber\\
\end{eqnarray}
Let us notice that $\bar{\omega}^{\mu}$ is related to the (normalized) thermal vorticity by $\omega^{\mu}=\zeta^{-1}\bar{\omega}^{\mu}$ with $\zeta\equiv\sqrt{-\bar{\omega}\cdot \bar{\omega}}$.
\par
In analogy to the method introduced in \cite{huang2022}, we use a decomposition of $\mu^{\mu}$ and $\sigma^{\mu}$ into the direction parallel and perpendicular to $\omega^{\mu}$ and $b^{\mu}$. The corresponding coefficients are denoted by $\mu_{a \|}$ and $\sigma_{a\|}$ with $a=\omega,b$  as well as $\mu_{\perp}^{\mu}$ and $\sigma_{\perp}^{\mu}$,
\begin{eqnarray}\label{N15a}
\mu^{\mu}&=&\mu_{\omega\|}  {\omega}^{\mu}+\mu_{b\|}b^{\mu}+\mu_{\perp}^{\mu},\nonumber\\
\sigma^{\mu}&=&\sigma_{\omega\|}  {\omega}^{\mu}+\sigma_{b\|}b^{\mu}+\sigma_{\perp}^{\mu}.
\end{eqnarray}
In this section, we assume $\omega\cdot b=0$. Hence, the subscript $\perp$ indicates the direction perpendicular to both $b^{\mu}$ and $\omega^{\mu}$. Using then the following power-counting scheme
\begin{eqnarray}\label{N16a}
&&\omega^{\mu}, b^{\mu}, u^{\mu},\mu_{b\|}, \mu_{\omega\|}, p_{i}, i=1,\cdots,6\sim \mathcal{O}(\partial^{0}),\nonumber\\
&&\sigma_{b\|}, \sigma_{\omega\|}\sim \mathcal{O}(\hbar), \nonumber\\
&&\delta\Theta^{\left(\mu\nu\right)}, \delta\Theta^{[\mu\nu]},\delta\tilde{F}^{\mu\nu}, \delta J^{\mu}, \mu^{\mu}_{\perp}\sim \mathcal{O}(\partial),\nonumber\\
&&\delta\sigma,\sigma_{\perp}^{\mu}\sim \mathcal{O}(\hbar\partial),
\end{eqnarray}
we arrive at
\begin{widetext}
\begin{eqnarray}\label{N17a}
\partial_{\mu}s^{\mu}&=&\theta\big[s-\beta\left(\epsilon+p_{1}-\mu n+MB\right)\big]+\big[
\left(p_2-\mu_{\omega\|}\sigma_{\omega\|}\right)  {\omega}^{\mu}  {\omega}^{\nu}+\left(p_4+B^2\left(1-\chi_m\right)-\mu_{b\|}\sigma_{b\|}\right)b^{\mu}b^{\nu}
\nonumber\\
&&+\left(p_6-\mu_{b\|}\sigma_{\omega\|}\right)b^{\mu}  {\omega}^{\nu}+\left(p_6-\mu_{\omega\|}\sigma_{b\|}\right)b^{\nu}  {\omega}^{\mu}\big]\partial_{\mu}\beta_{\nu}
+\left(p_{3}\omega^{\mu\nu}+p_{5}b^{\mu\nu}\right)\left(\partial_{\mu}\beta_{\nu}-\beta\mu_{\mu\nu}\right)\nonumber\\
&&+\partial_{\mu}\left(\beta\delta q^{\mu}+\beta \mu^{\mu} \delta\sigma+\delta s^{\mu}+\mu\beta\delta J^{\mu}-\beta M_{\nu}\delta\tilde{F}^{\mu\nu}\right)\nonumber\\
&&
-\delta J^{\mu}\partial_{\mu}\left(\beta\mu\right)
+\delta\Theta^{[\mu\nu]}\mathcal{A}_{\mu\nu}
+\delta\Theta^{(\mu\nu)}\mathcal{S}_{\mu\nu}+\delta\tilde{F}^{\mu\nu}\partial_{\mu}(\beta M_{\nu})
+\mathcal{O}\left(\hbar\partial^{2},\partial^{3}\right),
\end{eqnarray}

with
\begin{eqnarray}\label{N18a}
\mathcal{A}_{\mu\nu}\equiv\left(\partial_{[\mu}\beta_{\nu]}-\beta\mu_{\mu\nu}\right),\quad
\mathcal{S}_{\mu\nu}\equiv\partial_{(\mu}\beta_{\nu)}.
\end{eqnarray}
\end{widetext}
In the next section, we determine $p_{i}, i=1,\cdots,6$, as well as the dissipative parts of $J^{\mu}, \Theta^{\mu\nu},$ and $\tilde{F}^{\mu\nu}$ by making use of second law of thermodynamics.
\section{Dissipative SVMHD: Entropy current analysis}\label{sec2}
\subsection{Ideal SVMHD}\label{sec2B}
At this stage, we are in the position to determine $p_{i}, i=1,\cdots 6$ under the assumption $\omega\cdot b=0$. Here, we use of the second law of thermodynamics $\partial_{\mu}s^{\mu}= 0$ for the ideal fluid and set the coefficient of $\theta$ equal to zero. We arrive first at the Euler equation of thermodynamics
\begin{eqnarray}\label{N19a}
\epsilon+p_{0}=Ts+\mu n,
\end{eqnarray}
with $p_{0}\equiv p_{1}+MB$. This fixes $p_{1}$ as $p_{1}=p_{0}-MB$. Similarly, $p_{2}, p_{4},$ and $p_{6}$ are determined by setting the corresponding coefficients to $\omega^{\mu}\omega^{\nu}, b^{\mu}b^{\nu}$, and $ b^{\mu}\omega^{\nu}$ as well as $ b^{\nu}\omega^{\mu}$ equal to zero. We obtain
\begin{eqnarray}\label{N20a}
p_2&=&\mu_{\omega\|}\sigma_{\omega\|},\nonumber\\
p_4&=&-\left(1-\chi_m\right)B^2+\mu_{b\|}\sigma_{b\|},\nonumber\\
p_6&=&\mu_{b\|}\sigma_{\omega\|}=\mu_{\omega\|}\sigma_{b\|}.
\end{eqnarray}
Let us notice that $p_{6}$ is determined by making use of the assumption that the magnetic field and vorticity are perpendicular to each other (see Appendix \ref{appA} for more details).
According to this argument, $p_{3}$ and $p_{5}$ also vanish, because they have no counterparts in $\partial_{\mu}s^{\mu}$ from \eqref{N17a} and for $\omega\cdot b=0$ the bases $\omega^{\mu\nu}$ and $b^{\mu\nu}$ are independent. Plugging these results into \eqref{N7a}  and ignoring  the dissipative terms $\delta\Theta^{\mu\nu}$ and $\delta q^{\mu}$, we arrive at the energy-momentum tensor in the zeroth-order of derivative expansion for an ideal charged rotating and spinful fluid in the presence of a magnetic field, which is perpendicular to the vorticity of the fluid,
\begin{eqnarray}\label{N21a}
\Theta^{\mu\nu}_{0}&=&\epsilon u^{\mu}u^{\nu}-p_{\perp}\Xi^{\mu\nu}+p_{\times}  {\omega}^{\mu}  {\omega}^{\nu}+p_{\|}b^{\mu}b^{\nu}
\nonumber\\
&&+\frac{1}{2}B^2\left(u^{\mu}u^{\nu}-\Xi^{\mu\nu}-b^{\mu}b^{\nu}+  {\omega}^{\mu}  {\omega}^{\nu}\right)\nonumber\\
&&+2\mu_{b\|}\sigma_{\omega\|}b^{(\mu}  {\omega}^{\nu)},
\end{eqnarray}
where $p_{\perp}, p_{\times}, $ and $p_{\|}$ are defined by
\begin{eqnarray}\label{N22a}
p_{\perp}&\equiv& p_0-\chi_m B^2,\nonumber\\
p_{\times}&\equiv& p_{\perp}+\mu_{\omega\|}\sigma_{\omega\|},\nonumber\\
p_{\|}&\equiv& p_{0}+\mu_{b\|}\sigma_{b\|}.
\end{eqnarray}
Moreover, $\Xi^{\mu\nu}$ is defined by $\Xi^{\mu\nu}\equiv \Delta^{\mu\nu}+b^{\mu}b^{\nu}+ {\omega}^{\mu} {\omega}^{\nu}$.  In \eqref{N21a} and \eqref{N22a}, subscripts $\perp$ and $\|$ denote the perpendicular and parallel directions with respect to the direction of the magnetic field and vorticity. For vanishing $b^{\mu}$, \eqref{N22a} reduces to the same result presented in \cite{huang2022}. For vanishing $\omega^{\mu}$, the resulting energy-momentum tensor \eqref{N22a} coincides with the result presented in \cite{rischke2009,sadooghi2016}. The last term in \eqref{N21a} is a  novel term, proportional to $b^{(\mu}  {\omega}^{\nu)}$. It includes a coupling between the (thermal) vorticity and the magnetic field. The corresponding coefficient $\mu_{b\|}\sigma_{\omega\|}$, that, according to \eqref{N21a} is equal to $\mu_{\omega\|}\sigma_{b\|}$,  comprises the information from the spin chemical potential and spin.  In the absence of spin, this term thus vanishes. As it turns out, the magnetic susceptibility of the medium, $\chi_{m}$, which appears in $p_{\perp}$ and $p_{\times}$,  affects only the pressures which are perpendicular to the direction of the magnetic field. Moreover, whereas the cross pressure $p_{\times}$ consists of a term $\mu_{\omega\|}\sigma_{\omega\|}$, arising from the interaction of the vorticity and spin, the parallel pressure $p_{\|}$ includes a term $\mu_{b\|}\sigma_{b\|}$, arising from the interaction of the magnetic field and spin. The chemical potentials $\mu_{\omega\|}$ and $\mu_{b\|}$ can be interpreted as the amount of energy which is necessary to align the spin parallel to the vorticity $\bs{\omega}$ and the magnetic field $\bs{b}$, respectively.
\par
In what follows, we determine the complete
first-order constitutive relations of this model for the case $\omega\cdot b=0$.
\subsection{First order dissipative SVMHD: Transport coefficients}\label{sec2C}
Let us now focus on the dissipative part of $\partial_{\mu}s^{\mu}$ from \eqref{N17a}. From the term on the third line of \eqref{N17a}, we arrive at two expressions,
\begin{eqnarray}
\beta \mu^{\mu}\delta\sigma&=&-\beta\delta q^{\mu},\label{N23a}\\
\delta s^{\mu}&=&-\mu\beta\delta J^{\mu}+\beta M_{\nu}\delta\tilde{F}^{\mu\nu}. \label{N24a}
\end{eqnarray}
Plugging \eqref{N9a} into \eqref{N23a} and using $\beta\tilde{\omega}^{\mu}=\zeta\omega^{\mu}$, with $\tilde{\omega}^{\mu}=\frac{1}{2}\epsilon^{\mu\rho\nu\lambda}u_{\rho}\partial_{\nu}u_{\lambda}$, we arrive first at
\begin{eqnarray}\label{N25a}
\delta\sigma=\frac{\epsilon^{\mu\nu\rho\lambda}\beta u_{\rho}\partial_{\nu}\sigma_{\lambda}\left(\zeta\omega_{\mu}+\beta \mu_{\mu}\right)}{2\left(\zeta\omega_{\mu}+\beta\mu_{\mu}\right)\left(\zeta\omega^{\mu}+\beta\mu^{\mu}\right)}.
\end{eqnarray}
 Using then the decomposition \eqref{N15a} and the power-counting \eqref{N16a}, we obtain
\begin{eqnarray}\label{N26a}
\zeta{\omega}^{\mu}+\beta\mu^{\mu}&\approx&\left(\zeta+\beta\mu_{\omega\|}\right){\omega}^{\mu}+\beta\mu_{b\|}b^{\mu},\nonumber\\
\left(\zeta\omega_{\mu}+\beta\mu_{\mu}\right)\left(\zeta\omega^{\mu}+\beta\mu^{\mu}\right)&\approx&-\left(\zeta+\beta\mu_{\omega\|}\right)^{2}-\beta^{2}\mu_{b\|}^{2}, \nonumber\\
\end{eqnarray}
where we have used $b\cdot \omega=0$, as well as $b\cdot b=\omega\cdot\omega=-1$. The approximations in \eqref{N26a} are up to $\mathcal{O}(\partial)$. Finally, plugging \eqref{N26a} into \eqref{N23a}, $\delta \sigma$ is determined up to $\mathcal{O}\left(\hbar\partial^{2}\right)$,
\begin{eqnarray}\label{N27a}
\delta\sigma&=&\frac{\epsilon^{\mu\nu\rho\lambda}\beta u_{\rho}\partial_{\mu}\sigma_{\lambda}\{\left(\zeta+\beta\mu_{\omega\|}\right){\omega}_{\nu}+\beta\mu_{b\|}b_{\nu}\}}{2[\left(\zeta+\beta\mu_{\omega\|}\right)^{2}+\beta^{2}\mu_{b\|}^{2}]}\nonumber\\
&&+\mathcal{O}(\hbar\partial^2).
\end{eqnarray}
This confirms the assumption in \eqref{N16a}, according to which $\delta\sigma\sim \mathcal{O}(\hbar\partial)$.
\par
As concerns the second relation \eqref{N24a}, including $\delta J^{\mu}$ and $\delta\tilde{F}^{\mu\nu}$, as well as the symmetric and antisymmetric parts of $\Theta^{\mu\nu}$, appearing in \eqref{N17a}, they are determined by requiring $\partial_{\mu}s^{\mu}\geq 0$ and introducing the (thermal) conductivity tensor $\kappa^{\mu\nu}$, the viscous tensors $\gamma^{\mu\nu\rho\sigma}$ and $\eta^{\mu\nu\rho\sigma}$ as well as their cross-terms $\zeta^{\mu\nu\rho\sigma}$ and $\xi^{\mu\nu\rho\sigma}$,
 \begin{eqnarray}\label{N28a}
\delta J^{\mu}&=&-T\kappa^{\mu\nu}\partial_{\nu}\left(\beta\mu\right),\nonumber\\
\delta\Theta^{[\mu\nu]}&=&T\gamma^{\mu\nu\rho\sigma}\mathcal{A}_{\rho\sigma}+T\zeta^{\mu\nu\rho\sigma}\mathcal{S}_{\rho\sigma},\nonumber\\
\delta\Theta^{(\mu\nu)}&=&T\eta^{\mu\nu\rho\sigma}\mathcal{S}_{\rho\sigma}+T\xi^{\mu\nu\rho\sigma}\mathcal{A}_{\rho\sigma},
\end{eqnarray}
where $\mathcal{S}_{\rho\sigma}$ and $\mathcal{A}_{\rho\sigma}$ are defined in \eqref{N18a}.
In addition to these tensors, we have an additional coefficient $\rho^{\mu\nu\rho\sigma}$ in the decomposition of $\delta \tilde{F}^{\mu\nu}$, in analogy to \cite{hongo2022},
 \begin{eqnarray}\label{N29a}
\delta \tilde{F}^{\mu\nu}&=&T\rho^{\mu\nu\rho\sigma}\partial_{[\rho}\left(\beta M_{\sigma]}\right).
\end{eqnarray}
Let us notice that although the expression arising in \eqref{N28a} are the same as those appearing in \cite{huang2022}, because of the appearance a new basis corresponding to $b^{\mu}$ and, in particular, its combination to $\omega^{\mu}$, several additional transport coefficients appear in thermal conductivity, viscous, and electromagnetic resistivity tensors $\kappa^{\mu\nu},\gamma^{\mu\nu\rho\sigma},\eta^{\mu\nu\rho\sigma},\zeta^{\mu\nu\rho\sigma},\xi^{\mu\nu\rho\sigma}$, and $\rho^{\mu\nu\rho\sigma}$. In what follows, we determine these coefficients by choosing appropriate bases. Having in mind that in the Landau frame $u_{\mu}$ shall be orthogonal to the dissipative part of $\Theta^{\mu\nu}, J^{\mu},$ and $\tilde{F}^{\mu\nu},$ we have
\begin{eqnarray}\label{N30a}
&&u_{\mu}\kappa^{\mu\nu}=0,\quad
u_{\mu}\gamma^{\mu\nu\rho\sigma}=0,\quad
u_{\mu}\zeta^{\mu\nu\rho\sigma}=0,\nonumber\\
&&
u_{\mu}\eta^{\mu\nu\rho\sigma}=0, \quad u_{\mu}\xi^{\mu\nu\rho\sigma}=0, \quad u_{\mu}\rho^{\mu\nu\rho\sigma}=0.  \nonumber\\
\end{eqnarray}
Hence, the only appropriate primitive orthogonal bases are
\begin{eqnarray}\label{N31a}
b^{\mu},\qquad \omega^{\mu},\qquad \Xi^{\mu\nu}.
\end{eqnarray}
They satisfy $b_{\mu}\omega^{\mu}=b_{\mu}\Xi^{\mu\nu}=\omega_{\mu}\Xi^{\mu\nu}=0$. Defining the generalized form of the orthogonality relation by
$$(A\cdot B)^{\nu_{1}\cdots\nu_{m}}\equiv A_{\mu_{1}\cdots\mu_{n}}B^{\mu_{1}\cdots\mu_{n}\nu_{1}\cdots\nu_{m}}=0,$$
it turns out that $b_{\mu\nu}$ and $\omega_{\mu\nu}$ cannot be chosen as bases, because they are not orthogonal to the bases presented in \eqref{N31a}. Other important guidelines that are used to determine the above dissipative coefficients are listed below:
\begin{enumerate}

\item[i)] Onsager conditions:
\begin{eqnarray}\label{N34a}
\kappa^{\mu\nu}(\omega,B)&=&\kappa^{\nu\mu}(-\omega,-B),\nonumber\\
X^{\mu\nu\rho\sigma}(\omega,B)&=&X^{\rho\sigma\mu\nu}(-\omega,-B),
\end{eqnarray}
where $X=\{\eta,\zeta,\xi,\rho\}$. Moreover, we have
\begin{eqnarray}\label{N35a}
\zeta^{\mu\nu\rho\sigma}(\omega,B)=\xi^{\rho\sigma\mu\nu}(-\omega,-B).
\end{eqnarray}
\item[ii)] Because of symmetry arguments, $\eta^{\mu\nu\rho\sigma}$ has at least one symmetric part under $\rho\leftrightarrow\sigma$, which is symmetric under $\mu\leftrightarrow\nu$,
$\zeta^{\mu\nu\rho\sigma}$ has at least one symmetric part under $\rho\leftrightarrow\sigma$, which is antisymmetric under $\mu\leftrightarrow\nu$,
$\gamma^{\mu\nu\rho\sigma}$ and $\rho^{\mu\nu\rho\sigma}$ have at least one antisymmetric part under $\rho\leftrightarrow\sigma$, which are antisymmetric under $\mu\leftrightarrow\nu$, and
$\xi^{\mu\nu\rho\sigma}$ has at least one antisymmetric part under $\rho\leftrightarrow\sigma,$ which is symmetric under $\mu\leftrightarrow\nu$.
\end{enumerate}
To determine $\kappa^{\mu\nu}$, appearing in \eqref{N28a}, we use the ansatz,
\begin{eqnarray}\label{N36a}
\kappa^{\mu\nu} = \kappa_{\perp} \Xi^{\mu\nu} + \kappa_{{\omega}\|}  {\omega}^{\mu}  {\omega}^{\nu} + \kappa_{b\|} b^{\mu} b^{\nu} + 2\kappa_{\times} b^{(\mu} {\omega}^{\nu)},\nonumber\\
\end{eqnarray}
which is compatible with the above guidelines. Plugging the above expression into $\delta J^{\mu}$ from \eqref{N28a}, the resulting expression into \eqref{N17a} and using
\begin{eqnarray}\label{N37a}
\partial_{\mu}\left(\beta \mu\right)=\alpha_{\omega\|}\omega_{\mu}+\alpha_{b\|}b_{\mu}+\Xi_{\mu\beta}\partial^{\beta}(\beta\mu),
\end{eqnarray}
we arrive after some work at
\begin{eqnarray}\label{N38a}
-\delta J^{\mu}\partial_{\mu}\left(\beta \mu\right)&=&T\bigg[\kappa_{\perp}\left(\Xi^{\alpha\beta}\partial_{\beta}\left(\beta \mu\right)\right)^{2}\nonumber\\
&&+
\begin{pmatrix}
\alpha_{b\|}&\alpha_{\omega\|}
\end{pmatrix}
\begin{pmatrix}
\kappa_{b\|}&\kappa_{\times}\\
\kappa_{\times}&\kappa_{\omega\|}
\end{pmatrix}
\begin{pmatrix}
\alpha_{b\|}\\
\alpha_{\omega\|}
\end{pmatrix}
\bigg].\nonumber\\
\end{eqnarray}
This term appears in \eqref{N17a}. According to the second law of thermodynamics, $\partial_{\mu}s^{\mu}$ is to be positive definite. We thus have to determine appropriate conditions for the coefficients $\kappa_{\perp}, \kappa_{b\|}, \kappa_{\omega\|}$, and $\kappa_{\times}$ to guarantee the positive definiteness of \eqref{N38a}. These conditions are
\begin{eqnarray}\label{N39a}
\kappa_{\perp}\geq 0,\quad \kappa_{b\|}\geq 0,\quad \kappa_{b\|}\kappa_{\omega\|}-\kappa_{\times}^2\geq 0.
\end{eqnarray}
We further propose the following expressions for the dissipative coefficients $\gamma^{\rho\sigma\mu\nu}, \eta^{\rho\sigma\mu\nu}, \xi^{\rho\sigma\mu\nu}$, and $\rho^{\rho\sigma\mu\nu}$:
\begin{widetext}
\begin{eqnarray}\label{N40a}
	 \gamma^{\mu\nu\rho\sigma}&=& \gamma_{\perp}(\Xi^{\mu\rho}\Xi^{\nu\sigma}-\Xi^{\mu\sigma}\Xi^{\nu\rho}) + 2\gamma_{\omega\|}( {\omega}^{\mu}\Xi^{\nu[\rho} {\omega}^{\sigma]}- {\omega}^{\nu}\Xi^{\mu[\rho} {\omega}^{\sigma]}) + 2\gamma_{b\|}(b^{\mu}\Xi^{\nu[\rho}b^{\sigma]}-b^{\nu}\Xi^{\mu[\rho}b^{\sigma]})} + 4 \gamma_{\times}  {\omega}^{[\mu} b^{\nu]}  {\omega}^{[\rho} b^{\sigma]
	 \nonumber \\ &&
	+ 2\gamma_{\times}^{\prime}( {\omega}^{[\mu} \Xi^{\nu]\rho} b^{\sigma} -  {\omega}^{[\mu} \Xi^{\nu]\sigma} b^{\rho} + b^{[\mu} \Xi^{\nu]\rho}  {\omega}^{\sigma} - b^{[\mu} \Xi^{\nu]\sigma}  {\omega}^{\rho} ), \nonumber\\
\eta^{\mu\nu\rho\sigma} &=& \zeta_{\perp}\Xi^{\mu\nu}\Xi^{\rho\sigma}+\zeta_{\omega\|} {\omega}^{\mu} {\omega}^{\nu} {\omega}^{\rho} {\omega}^{\sigma}+\zeta_{b\|}b^{\mu}b^{\nu}b^{\rho}b^{\sigma}+\zeta_{\omega\times}( {\omega}^{\mu} {\omega}^{\nu}\Xi^{\rho\sigma}+ {\omega}^{\rho} {\omega}^{\sigma}\Xi^{\mu\nu})+ \zeta_{b\times}(b^{\mu}b^{\nu}\Xi^{\rho\sigma}+b^{\rho}b^{\sigma}\Xi^{\mu\nu}) \nonumber \\
&&
+2\zeta_{\times}^{(1)}(b^{(\mu} {\omega}^{\nu)} {\omega}^{\rho} {\omega}^{\sigma}+b^{(\rho} {\omega}^{\sigma)} {\omega}^{\mu} {\omega}^{\nu})
+2\zeta_{\times}^{(2)}( {\omega}^{(\mu}b^{\nu)}b^{\rho}b^{\sigma}+ {\omega}^{(\rho}b^{\sigma)}b^{\mu}b^{\nu})
+ 4\zeta_{\times} b^{(\mu}  {\omega}^{\nu)} b^{(\rho}  {\omega}^{\sigma)}\nonumber\\
&&
+\zeta_{\times}^{\prime}(b^{\mu}b^{\nu} {\omega}^{\rho} {\omega}^{\sigma}+b^{\rho}b^{\sigma} {\omega}^{\mu} {\omega}^{\nu})
 + 2\zeta_{\times}^{''} (\Xi^{\mu\nu} b^{(\rho}  {\omega}^{\sigma)} + \Xi^{\rho\sigma} b^{(\mu}  {\omega}^{\nu)} )  +\eta_{\perp}(\Xi^{\mu(\rho}\Xi^{\sigma)\nu}-\Xi^{\mu\nu}\Xi^{\rho\sigma})\nonumber \\
&&+2\eta_{\omega\|}( {\omega}^{\mu}\Xi^{\nu(\rho} {\omega}^{\sigma)}+ {\omega}^{\nu}\Xi^{\mu(\rho} {\omega}^{\sigma)})  +2\eta_{b\|}(b^{\mu}\Xi^{\nu(\rho}b^{\sigma)}+b^{\nu}\Xi^{\mu(\rho}b^{\sigma)}) + 4 \eta_{\times} ( {\omega}^{(\mu} \Xi^{\nu)(\rho} b^{\sigma)} + b^{(\mu} \Xi^{\nu)(\rho}  {\omega}^{\sigma)} ),\nonumber\\
\xi^{\mu\nu\rho\sigma}&=& 2 \xi_{b\|} (  {b}^{\mu} \Xi^{\nu[\rho}   {b}^{\sigma]} +   {b}^{\nu} \Xi^{\mu[\rho}   {b}^{\sigma]} +   {b}^{\nu} \Xi^{\mu(\rho}   {b}^{\sigma)} -   {b}^{\mu} \Xi^{\nu(\rho}   {b}^{\sigma)}) \nonumber \\ 
&& + 2 \xi_{\omega\|} (  {\omega}^{\mu} \Xi^{\nu[\rho}   {\omega}^{\sigma]} +   {\omega}^{\nu} \Xi^{\mu[\rho}   {\omega}^{\sigma]} +   {\omega}^{\nu} \Xi^{\mu(\rho}   {\omega}^{\sigma)} -   {\omega}^{\mu} \Xi^{\nu(\rho}   {\omega}^{\sigma)}) \nonumber \\ && + 2 \xi_{\times} (  {\omega}^{\mu}   {b}^{\nu}   {b}^{[\rho}   {\omega}^{\sigma]} +   {\omega}^{\nu}   {b}^{\mu}   {b}^{[\rho}   {\omega}^{\sigma]} +   {\omega}^{\nu}   {b}^{\mu}   {b}^{(\rho}   {\omega}^{\sigma)} -   {\omega}^{\mu}   {b}^{\nu}   {b}^{(\rho}   {\omega}^{\sigma)} ) \nonumber \\
&&  + 2 \xi^{\prime}_{\times} (  {b}^{\mu} \Xi^{\nu[\rho}   {\omega}^{\sigma]} +   {b}^{\nu} \Xi^{\mu[\rho}   {\omega}^{\sigma]} +   {\omega}^{\nu} \Xi^{\mu(\rho}   {b}^{\sigma)} -   {\omega}^{\mu} \Xi^{\nu(\rho}   {b}^{\sigma)}) \nonumber \\
&& + 2 \xi_{\times}^{''} (  {\omega}^{\mu} \Xi^{\nu[\rho}   {b}^{\sigma]} +   {\omega}^{\nu} \Xi^{\mu[\rho}   {b}^{\sigma]} +   {b}^{\nu} \Xi^{\mu(\rho}   {\omega}^{\sigma)} -   {b}^{\mu} \Xi^{\nu(\rho}   {\omega}^{\sigma)}) \nonumber \\
&& + 2 \zeta_{\times}^{(3)} (  {\omega}^{\mu}   {\omega}^{\nu}   {b}^{[\rho}   {\omega}^{\sigma]} +   {\omega}^{\rho}   {\omega}^{\sigma}   {b}^{[\mu}   {\omega}^{\nu]}) + 2 \zeta_{\times}^{(4)} (  {b}^{\mu}   {b}^{\nu}   {b}^{[\rho}   {\omega}^{\sigma]} +   {b}^{\rho}   {b}^{\sigma}   {b}^{[\mu}   {\omega}^{\nu]}) \nonumber \\
&& + 2 \zeta_{\times}^{(5)} (\Xi^{\mu\nu}   {\omega}^{[\rho}   {b}^{\sigma]} + \Xi^{\rho\sigma}   {\omega}^{[\mu}   {b}^{\nu]}),\nonumber\\
\rho^{\mu\nu\rho\sigma}&=& \rho_{\perp}(\Xi^{\mu\rho}\Xi^{\nu\sigma}-\Xi^{\mu\sigma}\Xi^{\nu\rho}) + 2\rho_{\omega\|}(  {\omega}^{\mu}\Xi^{\nu[\rho}  {\omega}^{\sigma]}-  {\omega}^{\nu}\Xi^{\mu[\rho}  {\omega}^{\sigma]}) + 2\rho_{b\|}(  {b}^{\mu}\Xi^{\nu[\rho}  {b}^{\sigma]}-  {b}^{\nu}\Xi^{\mu[\rho}  {b}^{\sigma]})} + 4 \rho_{\times}   {\omega}^{[\mu}   {b}^{\nu]}   {\omega}^{[\rho}   {b}^{\sigma]
	\nonumber \\ &&
	+ 2\rho_{\times}^{\prime}(  {\omega}^{[\mu} \Xi^{\nu]\rho}   {b}^{\sigma} -   {\omega}^{[\mu} \Xi^{\nu]\sigma}   {b}^{\rho} +   {b}^{[\mu} \Xi^{\nu]\rho}   {\omega}^{\sigma} -   {b}^{[\mu} \Xi^{\nu]\sigma}   {\omega}^{\rho} ).
\end{eqnarray}
\end{widetext}
The coefficient $\zeta^{\mu\nu\rho\sigma}$ is determined from $\xi^{\mu\nu\rho\sigma}$ by using \eqref{N35a}. At this stage, we have to insert the above expressions for the dissipative coefficients into \eqref{N28a} and the resulting expressions into $\partial_{\mu}s^{\mu}$. To satisfy the second law of thermodynamics, the final result shall be brought in a quadratic form. To do this, and, in particular, to find the conditions under which $\partial_{\mu}s^{\mu}\geq 0$, we go through the following steps. We first decompose $\mathcal{A}_{\mu\nu}$ and $\mathcal{S}_{\mu\nu}$, defined in \eqref{N18a} and appearing in \eqref{N17a} as well as \eqref{N28a}, in terms of appropriate bases,
\begin{eqnarray}\label{N41a}
	\mathcal{A}_{\mu\nu} &=& 2 \phi_{\times} b_{[\mu}  {\omega}_{\nu]} + 2 \left(
b_{[\mu} \mathcal{I}_{\nu]} +  {\omega}_{[\mu} \mathcal{I}_{\nu]}\right)  \nonumber\\
&& + (\Xi_{\mu\alpha} \Xi_{\nu\beta} - \Xi_{\mu\beta} \Xi_{\nu\alpha}) \mathcal{A}^{\alpha\beta},\nonumber\\
	\mathcal{S}_{\mu\nu} &=& \theta_{b\|} b_{\mu} b_{\nu} + \theta_{\perp} \Xi_{\mu\nu} + \theta_{\omega\|}  {\omega}_{\mu}  {\omega}_{\nu} + 2 \theta_{\times} b_{(\mu}  {\omega}_{\nu)} \nonumber \\
&&+ 2  \left(
b_{(\mu} \mathcal{J}_{\nu)} +  {\omega}_{(\mu} \mathcal{J}_{\nu)}\right)
+\partial_{\langle\mu}\beta_{\nu\rangle},
\end{eqnarray}
with
\begin{eqnarray}\label{N42a}
\mathcal{I}_{\rho}&\equiv& \left(b_{\alpha}+\omega_{\alpha}\right)\Xi_{\rho\beta}\mathcal{A}^{\alpha\beta},\nonumber\\
\mathcal{J}_{\rho}&\equiv& \left(b_{\alpha}+\omega_{\alpha}\right)\Xi_{\rho\beta}\mathcal{S}^{\alpha\beta},
\end{eqnarray}
and \cite{rischke-book}
\begin{eqnarray}\label{N43a}
\partial_{\langle\mu} \beta_{\nu\rangle} \equiv  \frac{1}{2} \left(\Xi_{\mu(\rho} \Xi_{\sigma)\nu} - \Xi_{\mu\nu} \Xi_{\rho\sigma}\right) \partial^{\rho} \beta^{\sigma}.
\end{eqnarray}
Plugging then \eqref{N40a} together with \eqref{N41a} into \eqref{N28a} and the resulting expressions into \eqref{N17a}, we arrive at
\begin{eqnarray}\label{N44a}
\Gamma_{2\rho}^{T}\boldsymbol{M}_{2}\Gamma_{2}^{\rho}+\Gamma_{5}^{T}\boldsymbol{M}_{5}\Gamma_{5},
\end{eqnarray}
with
\begin{eqnarray}\label{N45a}
\Gamma_{2\rho}^{T}&\equiv&
\begin{pmatrix}
\mathcal{J}_{\rho}&
\mathcal{I}_{\rho}
\end{pmatrix},\nonumber\\
\Gamma_{5}^{T}&\equiv&
\begin{pmatrix}
\theta_{\perp}&
\theta_{b \|}&
\theta_{\omega \|}&
\theta_{\times}&
\phi_{\times}
\end{pmatrix},
\end{eqnarray}
and
\begin{widetext}
\begin{eqnarray}\label{N46a}
\hspace{-0.5cm}\boldsymbol{M}_{2}=-4
\begin{pmatrix}
-\left(\eta_{\omega \|}+\eta_{b \|}+2\eta_{\times}\right)&
\xi_{\omega \|}+\xi_{b \|}+\xi^{\prime}_{\times}+\xi^{''}_{\times}\\
\xi_{\omega \|}+\xi_{b \|}+\xi^{\prime}_{\times}+\xi^{''}_{\times}&
\gamma_{\omega \|}+\gamma_{b \|}+2\gamma_{\times}^{\prime}
\end{pmatrix},
\quad
\boldsymbol{M}_{5}=
\begin{pmatrix}
\zeta_{\perp}&\zeta_{b\times}&\zeta_{\omega\times}&2\zeta^{''}_{\times}&-2\zeta_{\times}^{(5)}\\
\zeta_{b\times}&\zeta_{b \|}&\zeta_{\times}^{\prime}&2\zeta_{\times}^{(2)}&2\zeta_{\times}^{(4)}\\
\zeta_{\omega\times}&\zeta_{\times}^{\prime}&\zeta_{\omega \|}&2\zeta_{\times}^{(1)}&2\zeta_{\times}^{(3)}\\
2\zeta^{''}_{\times}&2\zeta_{\times}^{(2)}&2\zeta_{\times}^{(1)}&4\zeta_{\times}&4\xi_{\times}\\
-2\zeta_{\times}^{(5)}&2\zeta_{\times}^{(4)}&2\zeta_{\times}^{(3)}&4\xi_{\times}&4\gamma_{\times}
\end{pmatrix}.
\end{eqnarray}
The positivity of $\bs{M}_{2}$ and $\bs{M}_{5}$ guarantees the requirement $\partial_{\mu} s^{\mu}\geq 0$. Using a theorem from linear algebra \cite{glibert-strang}, according to which the (semi)positiveness of all upper $k\times k$ determinants of a symmetric $n\times n$ matrix guarantees the (semi)positiveness of this matrix, we obtain several conditions for the transport coefficients appearing in \eqref{N40a}.  Here, $k=1,\cdots, n$. From $\bs{M}_{2}$, we obtain two conditions,
\begin{eqnarray}\label{N47a}
\hspace{-1cm}\eta_{\omega \|}+\eta_{b \|}+2\eta_{\times}\geq 0,\qquad\mbox{and}\qquad \left(\eta_{\omega \|}+\eta_{b \|}+2\eta_{\times}\right)\left(\gamma_{\omega \|}+\gamma_{b \|}+2\gamma_{\times}^{\prime}\right)+\left(\xi_{\omega \|}+\xi_{b \|}+\xi^{\prime}_{\times}+\xi^{''}_{\times}\right)^{2}\leq 0.
\end{eqnarray}
Similarly,  $\bs{M}_{5}$ leads to five conditions,
\begin{eqnarray}\label{N48a}
&&\zeta_{\perp}\geq 0,\nonumber\\
&&\zeta_{\perp}\zeta_{b \|}-\zeta_{b\times}^{2}\geq 0,\nonumber\\
&&
\begin{vmatrix}
\zeta_{\perp}&\zeta_{b\times}&\zeta_{\omega\times}\\
\zeta_{b\times}&\zeta_{b \|}&\zeta_{\times}^{\prime}\\
\zeta_{\omega\times}&\zeta_{\times}^{\prime}&\zeta_{\omega \|}
\end{vmatrix}\geq 0,
\qquad
\begin{vmatrix}
\zeta_{\perp}&\zeta_{b\times}&\zeta_{\omega\times}&2\zeta^{''}_{\times}\\
\zeta_{b\times}&\zeta_{b \|}&\zeta_{\times}^{\prime}&2\zeta_{\times}^{(2)}\\
\zeta_{\omega\times}&\zeta_{\times}^{\prime}&\zeta_{\omega \|}&2\zeta_{\times}^{(1)}\\
2\zeta^{''}_{\times}&2\zeta_{\times}^{(2)}&2\zeta_{\times}^{(1)}&4\zeta_{\times}\\
\end{vmatrix}\geq 0,
\qquad
\begin{vmatrix}
\zeta_{\perp}&\zeta_{b\times}&\zeta_{\omega\times}&2\zeta^{''}_{\times}&-2\zeta_{\times}^{(5)}\\
\zeta_{b\times}&\zeta_{b \|}&\zeta_{\times}^{\prime}&2\zeta_{\times}^{(2)}&2\zeta_{\times}^{(4)}\\
\zeta_{\omega\times}&\zeta_{\times}^{\prime}&\zeta_{\omega \|}&2\zeta_{\times}^{(1)}&2\zeta_{\times}^{(3)}\\
2\zeta^{''}_{\times}&2\zeta_{\times}^{(2)}&2\zeta_{\times}^{(1)}&4\zeta_{\times}&4\xi_{\times}\\
-2\zeta_{\times}^{(5)}&2\zeta_{\times}^{(4)}&2\zeta_{\times}^{(3)}&4\xi_{\times}&4\gamma_{\times}
\end{vmatrix}\geq 0.
\end{eqnarray}
\end{widetext}
The last step is to determine $\rho^{\mu\nu\rho\sigma}$. Plugging the proposed expression for this coefficient from \eqref{N40a} into \eqref{N29a} and using
\begin{eqnarray}\label{N49a}
\lefteqn{\partial_{[\mu} (\beta M_{\nu]}) =2 \psi_{\times}  {b}_{[\mu}  {\omega}_{\nu]} + 2  {b}_{\alpha}  {b}_{[\mu} \Xi_{\nu]\beta} \partial^{[\alpha} (\beta M^{\beta]}) }
\nonumber\\
&&\quad+ 2  {\omega}_{\alpha}  {\omega}_{[\mu} \Xi_{\nu]\beta}  \partial^{[\alpha} (\beta M^{\beta]})  + 2  {b}_{\alpha}  {\omega}_{[\mu} \Xi_{\nu]\beta}  \partial^{[\alpha} (\beta M^{\beta]}) \nonumber\\
&&\quad + 2  {\omega}_{\alpha}  {b}_{[\mu} \Xi_{\nu]\beta}  \partial^{[\alpha} (\beta M^{\beta]}) \nonumber\\
&&\quad + (\Xi_{\mu\alpha} \Xi_{\nu\beta} - \Xi_{\mu\beta} \Xi_{\nu\alpha})  \partial^{[\alpha} (\beta M^{\beta]}),
\end{eqnarray}
we arrive at
\begin{eqnarray}\label{N50a}
\lefteqn{\hspace{-1cm}T \rho^{\mu\nu\rho\sigma} \partial_{[\mu} (\beta M_{\nu]}) \partial_{[\rho} (\beta M_{\sigma]}) =
}\nonumber\\
&& 8T \rho_{\perp}(\Xi_{\mu\gamma}\Xi_{\nu\delta}\partial^{[\gamma} (\beta M^{\delta]}))^2
+ 4T \rho_{\times} \psi_{\times}^{2}\nonumber\\
&&- 4T(\rho_{\omega\|} + \rho_{b\|} + 2 \rho_{\times}^{\prime}) \nonumber\\
&&\times [({b}_{\alpha} +  {\omega}_{\alpha}) \Xi_{\rho\beta} \partial^{[\alpha} (\beta M^{\beta]})]^2.
\end{eqnarray}
This leads immediately to the following conditions,
\begin{eqnarray}\label{N51a}
	\rho_{\perp} \geq 0, \;\;\; \rho_{\times} \geq 0, \;\;\; \rho_{\omega\|} + \rho_{b\|} +2 \rho_{\times}^{\prime} \leq 0,
\end{eqnarray}
that (partly) guarantee $\partial_{\mu}s^{\mu}\geq 0$. As a by-product, it is possible to determine the induced electric field $\delta E^{\mu}$ in the first-order derivative expansion from
\begin{eqnarray}\label{N52a}
	\delta E^{\mu} =  \frac{1}{2} \epsilon^{\mu\nu\alpha\beta} u_{\nu} \delta \tilde{F}_{\alpha\beta},
\end{eqnarray}
with $\delta\tilde{F}_{\alpha\beta}$ from \eqref{N29a} and the resistivity tensor $\rho^{\mu\nu\rho\sigma}$ from \eqref{N40a}. Moreover, $\delta s^{\mu}$ from \eqref{N24a} is also determined.
\par
In summary, the present SVMHD model includes $36$ transport coefficients in the first-order derivative expansion,
\begin{itemize}
\item[-] 4 thermal conductivities: $	\kappa_{\perp},  \kappa_{\omega\|},  \kappa_{b\|},  \kappa_{\times}, $
\item[-] 5 rotational viscosities: $\gamma_{\perp},  \gamma_{\omega\|},  \gamma_{b\|},  \gamma_{\times},  \gamma_{\times}^{\prime}, $
\item[-] 10 bulk viscosities: $\zeta_{\perp},  \zeta_{\omega\|},  \zeta_{b\|},  \zeta_{\omega\times},  \zeta_{b\times},  \zeta_{\times}^{(1)},$
$\zeta_{\times}^{(2)},  \zeta_{\times},  \zeta_{\times}^{\prime},  \zeta_{\times}^{''},$
\item[-] 4 shear viscosities: $\eta_{\perp},  \eta_{\omega\|},  \eta_{b\|},  \eta_{\times},$
\item[-] 8 cross viscosities: $\xi_{b\|},  \xi_{\omega\|},  \xi_{\times},  \xi_{\times}^{\prime},  \xi_{\times}^{''},  \zeta_{\times}^{(3)},\zeta_{\times}^{(4)},$  $\zeta_{\times}^{(5)},$
\item[-] 5 electric resistivities: $
\rho_{\perp},  \rho_{\omega\|},  \rho_{b\|},  \rho_{\times},   \rho_{\times}^{\prime}.$
\end{itemize}
The second law of thermodynamics lead to the conditions \eqref{N47a}, \eqref{N48a}, and \eqref{N51a} for the above coefficients. In contrast to the results presented in \cite{huang2022}, for the case where the magnetic field is absent but the vorticity is present, no Hall coefficients appear in the list of the transport coefficients presented above. We finally notice that the traceless bases correspond only to rotational, shear, and electric viscosities as well as to cross viscosities $\xi_{b\|}, \xi_{\omega\|},\xi_{\times},\xi_{\times}^{\prime},$ and $\xi_{\times}^{''}$. All the other bases are traceful.
\subsection{A possible modification of SVMHD: Working with electric part of the thermal vorticity}\label{sec2D}
In the first part of this section, we adopt the definition of the thermal vorticity tensor $\omega^{\mu\nu}=\epsilon^{\mu\nu\rho\sigma}\omega_{\rho}u_{\sigma}$ from \cite{huang2022}. But, as it is argued in e.g. \cite{florkowski2017}, the general definition of the rank-two tensor $\omega^{\mu\nu}$ is given by
\begin{eqnarray}\label{N53a}
\omega^{\mu\nu}=\kappa^{\mu}u^{\nu}-\kappa^{\nu}u^{\mu}-\epsilon^{\mu\nu\rho\sigma}\omega_{\rho}u_{\sigma}.
\end{eqnarray}
It includes an electric part, $\kappa^{\mu}u^{\nu}-\kappa^{\nu}u^{\mu}$ and a magnetic part,
$-\epsilon^{\mu\nu\rho\sigma}\omega_{\rho}u_{\sigma}$, in analogy to the definition of the field strength tensor $F^{\mu\nu}$,
\begin{eqnarray}\label{N54a}
F^{\mu\nu}=E^{\mu}u^{\nu}-E^{\nu}u^{\mu}-\epsilon^{\mu\nu\rho\sigma}B_{\rho}u_{\sigma}.
\end{eqnarray}
Hence, the vector $\kappa^{\mu}$ in the electric part of the vorticity is the electric vorticity tensor vector, whereas $\omega^{\mu}$ in the magnetic part of the vorticity plays the role of the magnetic vorticity vector.
\par
Repeating the analysis of the previous section by replacing $\omega^{\mu\nu}=\epsilon^{\mu\nu\rho\sigma}\omega_{\rho}u_{\sigma}$ with $\omega^{\mu\nu}=\kappa^{\mu}u^{\nu}-\kappa^{\nu}u^{\mu}$, and $\omega^{\mu}$ with $\kappa^{\mu}$, we arrive first at the energy-momentum tensor for an ideal  fluid in the presence of external magnetic field and electric vorticity,
\begin{eqnarray}\label{N55a}
\Theta^{\mu\nu}_{0}&=&\epsilon u^{\mu}u^{\nu}-p_{\perp}\Xi^{\mu\nu}+p_{\times}  {\kappa}^{\mu}  {\kappa}^{\nu}+p_{\|}b^{\mu}b^{\nu}
\nonumber\\
&&+\frac{1}{2}B^2\left(u^{\mu}u^{\nu}-\Xi^{\mu\nu}-b^{\mu}b^{\nu}+  {\kappa}^{\mu}  {\kappa}^{\nu}\right)\nonumber\\
&&+2\mu_{b\|}\sigma_{k\|}b^{(\mu}  {\kappa}^{\nu)},
\end{eqnarray}
where $p_{\perp}, p_{\times}$, and $p_{\|}$ are given in \eqref{N22a}, and $\Xi^{\mu\nu}$ is defined by $\Xi^{\mu\nu}\equiv \Delta^{\mu\nu}+b^{\mu}b^{\nu}+ {\kappa}^{\mu} {\kappa}^{\nu}$, in analogy to $\Xi^{\mu\nu}$ defined in the previous section.
\par
As concerns the dissipative parts of $J^{\mu},\Theta^{\mu\nu}$, and $\tilde{F}^{\mu\nu}$, they  are determined by using the decomposition \eqref{N28a} and the following ans\"atze for the corresponding dissipative coefficients,
\begin{widetext}
\begin{eqnarray}\label{N56a}
\kappa^{\mu\nu} &=& \kappa_{\perp} \Xi^{\mu\nu} + \kappa_{{b\|}} {b}^{\mu} {b}^{\nu}  + \kappa_{{\kappa\|}} {\kappa}^{\mu} {\kappa}^{\nu} + 2\kappa_{{H\times}} {b}^{[\mu} {\kappa}^{\nu]},\nonumber\\
\gamma^{\mu\nu\rho\sigma} &=& \gamma_{\perp} (\Xi^{\mu\rho} \Xi^{\nu\sigma} - \Xi^{\mu\sigma} \Xi^{\nu\rho}) + 2\gamma_{{\kappa\|}} ( {\kappa}^{\mu} \Xi^{\nu[\rho} {\kappa}^{\sigma]} - {\kappa}^{\nu} \Xi^{\mu[\rho} {\kappa}^{\sigma]})  + 2\gamma_{{b\|}} ( {b}^{\mu} \Xi^{\nu[\rho} {b}^{\sigma]} - {b}^{\nu} \Xi^{\mu[\rho} {b}^{\sigma]}) + 4 \gamma_{\times} {\kappa}^{[\mu} {b}^{\nu]} {\kappa}^{[\rho} {b}^{\sigma]} \nonumber \\
&& + 2\gamma_{{H\times}}({\kappa}^{[\mu} \Xi^{\nu]\rho} {b}^{\sigma} - {\kappa}^{[\mu} \Xi^{\nu]\sigma} {b}^{\rho} - {b}^{[\mu} \Xi^{\nu]\rho} {\kappa}^{\sigma} + {b}^{[\mu} \Xi^{\nu]\sigma} {\kappa}^{\rho} ), \nonumber \\
\eta^{\mu\nu\rho\sigma} &=& \zeta_{\perp} \Xi^{\mu\nu} \Xi^{\rho \sigma} + \zeta_{{\kappa\|}} {\kappa}^{\mu} {\kappa}^{\nu} {\kappa}^{\rho} {\kappa}^{\sigma} + \zeta_{{b\|}} {b}^{\mu} {b}^{\nu} {b}^{\rho} {b}^{\sigma} + \zeta_{{\kappa\times}} ({\kappa}^{\mu} {\kappa}^{\nu} \Xi^{\rho\sigma} + {\kappa}^{\rho} {\kappa}^{\sigma} \Xi^{\mu \nu}) \nonumber \\
&& + \zeta_{{b\times}} ({b}^{\mu} {b}^{\nu} \Xi^{\rho\sigma} + {b}^{\rho} {b}^{\sigma} \Xi^{\mu \nu} ) + 4\zeta_{\times} ({b}^{(\mu} {\kappa}^{\nu)} {b}^{(\rho} {\kappa}^{\sigma)}) + \zeta^{\prime}_{\times} ({b}^{\mu} {b}^{\nu} {\kappa}^{\rho} {\kappa}^{\sigma} + {b}^{\rho} {b}^{\sigma} {\kappa}^{\mu} {\kappa}^{\nu}) \nonumber \\
&& + 2\zeta_{{H\times}} (\Xi^{\mu\nu} {b}^{(\rho} {\kappa}^{\sigma)} - \Xi^{\rho\sigma} {b}^{(\mu} {\kappa}^{\nu)} ) + 2\zeta_{{H}_{\times}}^{\prime} ({b}^{(\mu}{\kappa}^{\nu)} {\kappa}^{\rho} {\kappa}^{\sigma} - {b}^{(\rho} {\kappa}^{\sigma)} {\kappa}^{\mu} {\kappa}^{\nu} ) + 2\zeta_{{H}_{\times}}^{''} ({\kappa}^{(\mu}{b}^{\nu)} {b}^{\rho} {b}^{\sigma} - {\kappa}^{(\rho} {b}^{\sigma)} {b}^{\mu} {b}^{\nu}) \nonumber \\
&& + \eta_{\perp} (\Xi^{\mu(\rho} \Xi^{\sigma)\nu} - \Xi^{\mu\nu} \Xi^{\rho\sigma} )+ 2\eta_{{\kappa\|}} ( {\kappa}^{\mu} \Xi^{\nu(\rho} {\kappa}^{\sigma)} + {\kappa}^{\nu} \Xi^{\mu(\rho} {\kappa}^{\sigma)} ) + 2\eta_{{b\|}} ( {b}^{\mu} \Xi^{\nu(\rho} {b}^{\sigma)} + {b}^{\nu} \Xi^{\mu(\rho} {b}^{\sigma)} ) \nonumber \\
&&  + 2 \eta_{H} ( \Xi^{\mu\rho} {b}^{[\nu} {\kappa}^{\sigma]} + \Xi^{\nu \sigma} {b}^{[\mu} {\kappa}^{\rho]} + \Xi^{\mu \sigma} {b}^{[\nu} {\kappa}^{\rho]} + \Xi^{\nu\rho} {b}^{[\mu} {\kappa}^{\sigma]} ),  \nonumber \\
\xi^{\mu\nu\rho\sigma} &=& 2 \xi_{{b\|}} ( {b}^{\mu} \Xi^{\nu[\rho} {b}^{\sigma]} + {b}^{\nu} \Xi^{\mu[\rho} {b}^{\sigma]} + {b}^{\nu} \Xi^{\mu(\rho} {b}^{\sigma)} - {b}^{\mu} \Xi^{\nu(\rho} {b}^{\sigma)}) \nonumber \\
 &&+  2 \xi_{{\kappa\|}} ( {\kappa}^{\mu} \Xi^{\nu[\rho} {\kappa}^{\sigma]} + {\kappa}^{\nu} \Xi^{\mu[\rho} {\kappa}^{\sigma]} + {\kappa}^{\nu} \Xi^{\mu(\rho} {\kappa}^{\sigma)} - {\kappa}^{\mu} \Xi^{\nu(\rho} {\kappa}^{\sigma)}) \nonumber \\
&& + 2 \xi_{\times} ({\kappa}^{\mu} {b}^{\nu} {b}^{[\rho} {\kappa}^{\sigma]} + {\kappa}^{\nu} {b}^{\mu} {b}^{[\rho} {\kappa}^{\sigma]} + {\kappa}^{\nu} {b}^{\mu} {b}^{(\rho} {\kappa}^{\sigma)} - {\kappa}^{\mu} {b}^{\nu} {b}^{(\rho} {\kappa}^{\sigma)} )  \nonumber \\
&& + 2\xi_{H\times} ({\kappa}^{\mu} \Xi^{\nu[\rho} {b}^{\sigma]} + {\kappa}^{\nu} \Xi^{\mu[\rho} {b}^{\sigma]} -{b}^{\nu} \Xi^{\mu(\rho} {\kappa}^{\sigma)} + {b}^{\mu} \Xi^{\nu(\rho} {\kappa}^{\sigma)}) \nonumber \\
&& + 2\xi_{H\times}^{\prime} ( {b}^{\mu} \Xi^{\nu[\rho} {\kappa}^{\sigma]} + {b}^{\nu} \Xi^{\mu[\rho} {\kappa}^{\sigma]} - {\kappa}^{\nu} \Xi^{\mu(\rho} {b}^{\sigma)}+ {\kappa}^{\mu} \Xi^{\nu(\rho} {b}^{\sigma)}) \nonumber \\
&& + 2 \zeta_{{H}_{\times}}^{(1)} ( {\kappa}^{\mu} {\kappa}^{\nu} {\kappa}^{[\rho} {b}^{\sigma]} - {\kappa}^{\rho} {\kappa}^{\sigma} {\kappa}^{[\mu} {b}^{\nu]}) + 2 \zeta_{{H}_{\times}}^{(2)} ( {b}^{\mu} {b}^{\nu} {b}^{[\rho} {\kappa}^{\sigma]} - {b}^{\rho} {b}^{\sigma} {b}^{[\mu} {\kappa}^{\nu]})+ 2 \zeta_{H\times}^{(3)} (\Xi^{\mu\nu} {\kappa}^{[\rho} {b}^{\sigma]} - \Xi^{\rho\sigma} {\kappa}^{[\mu} {b}^{\nu]}),\nonumber\\
\rho^{\mu\nu\rho\sigma} &=& \rho_{\perp} (\Xi^{\mu\rho} \Xi^{\nu\sigma} - \Xi^{\mu\sigma} \Xi^{\nu\rho}) + 2\rho_{{\kappa\|}} ( {\kappa}^{\mu} \Xi^{\nu[\rho} {\kappa}^{\sigma]} - {\kappa}^{\nu} \Xi^{\mu[\rho} {\kappa}^{\sigma]})  + 2\rho_{{b\|}} ( {b}^{\mu} \Xi^{\nu[\rho} {b}^{\sigma]} - {b}^{\nu} \Xi^{\mu[\rho} {b}^{\sigma]}) + 4 \rho_{\times} {\kappa}^{[\mu} {b}^{\nu]} {\kappa}^{[\rho} {b}^{\sigma]} \nonumber \\
&& + 2\rho_{{H\times}}({\kappa}^{[\mu} \Xi^{\nu]\rho} {b}^{\sigma} - {\kappa}^{[\mu} \Xi^{\nu]\sigma} {b}^{\rho} - {b}^{[\mu} \Xi^{\nu]\rho} {\kappa}^{\sigma} + {b}^{[\mu} \Xi^{\nu]\sigma} {\kappa}^{\rho} ).
\end{eqnarray}
\end{widetext}
Similar to the previous case, $\zeta^{\mu\nu\rho\sigma}$ is determined by using \eqref{N35a}.
In contrast to the previous case, we have apart from 25 dissipative coefficients, including
\begin{itemize}
\item[-] $3$ thermal conductivities: $	\kappa_{\perp},  \kappa_{\kappa\|},  \kappa_{b\|},$
\item[-] $4$ rotational viscosities: $	\gamma_{\perp},  \gamma_{\kappa\|},  \gamma_{b\|},  \gamma_{\times},$
\item[-] $7$ bulk viscosities: $
	\zeta_{\perp},  \zeta_{\kappa\|},  \zeta_{b\|},  \zeta_{\kappa\times},\zeta_{b\times},   \zeta_{\times},$
$\zeta_{\times}^{\prime},$
\item[-] $3$ shear viscosities: $
\eta_{\perp},  \eta_{\kappa\|},\eta_{b\|},$
\item[-] $3$ cross viscosities: $	\xi_{b\|},\xi_{\kappa\|},\xi_{\times},$
\item[-] $4$ electric resistivities: $
	\rho_{\perp},\rho_{\kappa\|},\rho_{b\|},\rho_{\times},$
\end{itemize}
$11$ (nondissipative) Hall coefficients, including
\begin{itemize}
\item[-] $1$ thermal Hall conductivity: $	\kappa_{H\times},$
\item[-] $1$ rotational Hall viscosity: $\gamma_{H\times},$
\item[-] $3$ bulk Hall viscosities: $\zeta_{H\times},\zeta^{\prime}_{H\times},\zeta^{''}_{H\times},$
\item[-] $1$ shear Hall viscosity: $\eta_{H},$
\item[-] $5$ cross Hall viscosities: $\xi_{H\times},\xi^{\prime}_{H\times},\zeta^{(1)}_{H\times},$ $\zeta^{(2)}_{H\times},\zeta^{(3)}_{H\times},$
\item[-] $1$ electric Hall resistivity: $	\rho_{H\times} .$
\end{itemize}
In contrast to our results from previous section, a large number of Hall coefficients appear in this modified SVMHD model. This is mainly the consequence of different Onsager conditions in comparison with \eqref{N34a}. Having in mind that, in contrast to the magnetic spin vector $\omega^{\mu}$, the electric spin vector $\kappa^{\mu}$ is even under time reversal, the corresponding Onsager relations in the modified SVMHD are
\begin{eqnarray}\label{N57a}
\kappa^{\mu\nu}(\kappa,B)&=&\kappa^{\nu\mu}(\kappa,-B),\nonumber\\
X^{\mu\nu\rho\sigma}(\kappa,B)&=&X^{\rho\sigma\mu\nu}(\kappa,-B),
\end{eqnarray}
with $X=\{\eta,\zeta,\xi,\rho\}$.
Moreover, we have
\begin{eqnarray}\label{N58a}
\zeta^{\mu\nu\rho\sigma}(\kappa,B)=\xi^{\rho\sigma\mu\nu}(\kappa,-B).
\end{eqnarray}
 Following the same steps as described in the previous section, we arrive by  requiring $\partial_{\mu}s^{\mu}\geq 0$ at the following constraints for the dissipative coefficients:
\begin{eqnarray}\label{N59a}
&& \kappa_{\perp} \ge 0 , \; \kappa_{b\|} \ge 0 , \; \kappa_{\kappa\|} \ge 0, \nonumber\\
&& \eta_{\perp} \ge 0, \;\; \gamma_{\perp} \ge 0 , \; \; \eta_{\kappa\|} + \eta_{b\|} \ge 0,  \nonumber \\
&&  (\eta_{\kappa\|} + \eta_{b\|})( \gamma_{\kappa\|} + \gamma_{b\|}) + (\xi_{\kappa\|} + \xi_{b\|})^{2} \leq 0,  \nonumber \\
&& \zeta_{\perp} \ge 0 , \; \; \zeta_{\kappa\|} \ge 0, \; \; \zeta_{b\|} \ge 0, \; \; \zeta_{\times} \ge 0, \nonumber \\
&&  \zeta_{\perp} \zeta_{b\|} - \zeta_{{\times}_{b}}^{2} \ge 0, \; \; \zeta_{\perp} \zeta_{\kappa\|} - \zeta_{{\times}_{\kappa}}^{2} \ge, \; \; \zeta_{b\|} \zeta_{\kappa\|} - \zeta_{\times}^{\prime 2} \ge 0, \nonumber \\
&& \zeta_{b\|} (\zeta_{\perp} \zeta_{\kappa\|} - \zeta_{{\times}_{\kappa}}^{2}) + \zeta_{\kappa\|} ( \zeta_{\perp} \zeta_{b\|} - \zeta_{{\times}_{b}}^{2}) \nonumber\\
&&\qquad
+ \zeta_{\perp} ( \zeta_{b\|} \zeta_{\kappa\|} - \zeta_{\times}^{\prime 2}) \ge 2(-\zeta_{b\times}\zeta_{\kappa\times}\zeta_{\times}^{\prime} + \zeta_{\perp} \zeta_{b\|} \zeta_{\kappa\|}), \nonumber \\
&&\gamma_{\times} \zeta_{\times} - \xi_{\times}^2\geq 0,\nonumber\\
&& 	\rho_{\perp} \geq 0, \;\;\; \rho_{\times} \geq 0, \;\;\; \rho_{\kappa\|} + \rho_{b\|}  \leq 0.
\end{eqnarray}
It is worth to note that there is no constraint on the (nondissipative) Hall coefficients because they do not appear in $\partial_{\mu}s^{\mu}$ from \eqref{N17a}.
\section{Concluding remarks}\label{sec4}
In this paper, we generalized the recently introduced formulation of gyrohydrodynamics \cite{huang2022} for a spinful and vortical relativistic fluid by introducing an external magnetic field. We referred to the resulting framework  as the relativistic SVMHD. We aimed to study the interplay between the vorticity, magnetic field, and spin. As it is argued in \cite{huang2022}, the latter is, in contrast to the former, a quantum mechanical object and must be treated consequently. After introducing our model in Sec. \ref{sec2A}, we performed in Sec. \ref{sec2} an entropy current analysis by extending the method introduced in \cite{huang2022} to MHD. We determined, in particular, the energy-momentum tensor for an ideal SVMHD in terms of three different pressures $p_{\perp}, p_{\times},$ and $p_{\|}$ [see \eqref{N21a} for the case $\omega\cdot b=0$ and \eqref{appA9} for the case when $\omega\cdot b=\pm 1$]. According to their definitions, the magnetic susceptibility of the medium, which appears in $p_{\perp}$ and $p_{\times}$, affects only the pressures perpendicular to the direction of the magnetic field. It turned out that the cross pressure $p_{\times}$ consists of a term $\mu_{\omega\|}\sigma_{\omega\|}$, arising from the interaction of the vorticity and spin, the parallel pressure $p_{\|}$ includes a term $\mu_{b\|}\sigma_{b\|}$, arising from the interaction of the magnetic field and spin. Finally, the interaction of the magnetic field and the vorticity is reflected in a mixed term, proportional to $\mu_{b\|}\sigma_{\omega\|}$.
\par
In Sec. \ref{sec2C}, we determined the first-order constitutive relations of SVMHD by extending the power counting in \cite{huang2022} to the case of nonvanishing magnetic fields. We found that SVMHD consists of $36$ transport coefficients. In contrast to the case of zero magnetic fields, discussed in \cite{huang2022}, all coefficients contribute to entropy production. This effect is known from the literature and occurs when magnetic fields are introduced in hydrodynamics (for a recent review of MHD, see\cite{hongo2020}). In Sec. \ref{sec2C}, we determined the constraints (inequalities) these coefficients must satisfy in order to guarantee the positiveness of the entropy production rates.
\par
In this standard SVMHD formulation, we used the magnetic part of (thermal) vorticity $\omega^{\mu\nu}= \epsilon^{\mu\nu\rho\sigma}u_{\rho}\omega_{\sigma}$ [see \eqref{N51a}]. A reformulation of SVMHD in which the electric part of $\omega^{\mu\nu}$, i.e., $\omega^{\mu\nu}=\kappa^{\mu}u^{\nu}-\kappa^{\nu}u^{\mu}$ replaces the magnetic part of $\omega^{\mu\nu}$, i.e. $\omega^{\mu\nu}= \epsilon^{\mu\nu\rho\sigma}u_{\rho}\omega_{\sigma}$ leads, in contrast to the previous case, to $11$ nondissipative Hall-like coefficients, apart from $25$ dissipative ones. This difference arises from different behavior of the electric and magnetic part of the thermal vorticity under time-reversal transformation.  In Sec. \ref{sec2D}, the complete first-order constitutive relations for this modified SVMHD are presented. We showed that the inequalities that the dissipative transport coefficients satisfy in the modified case have a much simpler form than the original formulation of SVMHD with
$\omega^{\mu\nu}= \epsilon^{\mu\nu\rho\sigma}u_{\rho}\omega_{\sigma}$ from Sec. \ref{sec2C}.
\par
 It would be intriguing to use the results presented in the present paper to study the stability and causality of waves propagating in a dissipative spinful and vortical fluid in the presence of external magnetic fields. The stability and causality of anomalous MHD and spin hydrodynamics are studied in
\cite{shokri2021} and \cite{ hattori2019,mishra2022,ryblewski2022, nora2023, pu2023}, respectively. It would be important to extend these works to SVMHD and elucidate, in particular, the role played by the mixed term including $\omega^{\mu}$ and $b^{\mu}$, appearing in the energy-momentum tensor of ideal SVMHD. We notice that recently in \cite{shokri2023}, a novel method is introduced to determine local plane-wave solutions of the linearized hydrodynamic equation describing a fluid that rigidly moves with nonzero thermal vorticity. It would be interesting to use this method to determine the linear modes in the model introduced in the present paper.  We postpone this work to our future publications.
\begin{appendix}
\section{Energy-momentum tensor of ideal SVMHD: General case}\label{appA}
\setcounter{equation}{0}
In Sec. \ref{sec2B}, we derived the energy-momentum tensor of ideal SVMHD with $\omega\cdot b=0$. To do this, we used \eqref{N17a}, and the assumption $\omega\cdot b=0$. We argued that $p_{3}$ and $p_{5}$ vanish and $p_{2},p_{4},$ and $p_{6}$ are given by \eqref{N20a}. Plugging these expressions into $\Theta^{\mu\nu}$ from \eqref{N7a}, we arrived at \eqref{N21a} for the case when $\omega\cdot b=0$ (the vorticity and magnetic field are perpendicular). In this appendix, we present a more systematic method to determine $\Theta^{\mu\nu}$ for a general case $\omega\cdot b\neq 0$.
\par
Let us consider \eqref{N17a}. Using $\partial_{\mu}s^{\mu}=0$, as in the case $\omega\cdot b=0$, $p_{1}=p_{0}-MB$. Hence, \eqref{N19a} is still valid. For $p_{2}, p_{4}, p_{3}$, and $p_{5}$, we use the energy-momentum tensor in two different cases $\omega^{\mu}\neq 0\wedge b^{\mu}=0$ and $\omega^{\mu}= 0\wedge b^{\mu}\neq 0$, separately. According to \cite{huang2022}, when the magnetic field is absent and only a finite (thermal) vorticity is present, $p_{2}=\mu_{\omega\|}\sigma_{\omega\|}$ and $p_{3}=0$. This gives rise to $p_{4}=-B^{2}(1-\chi_{m}-\mu_{b\|}\sigma_{b\|})$ and $p_{5}=0$. The former reduces to $p_{4}=-B^{2}(1-\chi_{m})$, once the fluid is spinless. This term is expected from MHD.
\par
To determine $p_{6}$, we multiply the expression
\begin{eqnarray}\label{appA1}
\left(p_6-\mu_{b\|}\sigma_{\omega\|}\right)b^{\mu}  {\omega}^{\nu}+\left(p_6-\mu_{\omega\|}\sigma_{b\|}\right)b^{\nu}  {\omega}^{\mu},
\end{eqnarray}
with $\omega_{\mu}\omega_{\nu}, b_{\mu}b_{\nu},b_{\mu}\omega_{\nu},$ and $b_{\nu}\omega_{\mu}$ and set the resulting expression equal to zero. Multiplying \eqref{appA1} with  $\omega_{\mu}\omega_{\nu}$ and $b_{\mu}b_{\nu}$, and using $\omega\cdot\omega=-1$ as well as $b\cdot b=-1$ leads to
\begin{eqnarray}\label{appA2}
(\omega\cdot b)\left(2p_6-\mu_{b\|}\sigma_{\omega\|}-\mu_{\omega\|}\sigma_{b\|}\right)=0.
\end{eqnarray}
Multiplying \eqref{appA1} with $b_{\mu}\omega_{\nu}$ leads to
\begin{eqnarray}\label{appA3}
\left(p_6-\mu_{b\|}\sigma_{\omega\|}\right)+\left(\omega\cdot b\right)^{2}\left(p_6-\mu_{\omega\|}\sigma_{b\|}\right)=0.
\end{eqnarray}
Similarly, multiplying \eqref{appA1} with $b_{\nu}\omega_{\mu}$ yields
\begin{eqnarray}\label{appA4}
\left(\omega\cdot b\right)^{2}\left(p_6-\mu_{b\|}\sigma_{\omega\|}\right)+\left(p_6-\mu_{\omega\|}\sigma_{b\|}\right)=0.
\end{eqnarray}
Combining \eqref{appA3} and \eqref{appA4}, we arrive at
\begin{eqnarray}\label{appA5}
\lefteqn{\hspace{-1.5cm}[1-(\omega\cdot b)^{2}]\left(p_6-\mu_{b\|}\sigma_{\omega\|}\right)}\nonumber\\
&&=[1-(\omega\cdot b)^{2}]\left(p_6-\mu_{\omega\|}\sigma_{b\|}\right).
\end{eqnarray}
For $\left(\omega\cdot b\right)^2\neq 1$, \eqref{appA5} leads to
\begin{eqnarray}\label{appA6}
\mu_{b\|}\sigma_{\omega\|}=\mu_{\omega\|}\sigma_{b\|}.
\end{eqnarray}
Plugging \eqref{appA6} into \eqref{appA3} or \eqref{appA4}, we immediately arrive at
\begin{eqnarray}\label{appA7}
\hspace{-0.5cm}p_{6}=\mu_{b\|}\sigma_{\omega\|}=\mu_{\omega\|}\sigma_{b\|}, \quad\mbox{for}\quad (\omega\cdot b)^{2}\neq 1.
\end{eqnarray}
For $(\omega\cdot b)^{2}=1$, however, \eqref{appA5} is trivially satisfied. To determine the most general form of $p_{6}$ is this case, we have to use \eqref{appA2}. The resulting expression reads
\begin{eqnarray}\label{appA8}
\hspace{-0.5cm}p_{6}=\frac{1}{2}\left(\mu_{b\|}\sigma_{\omega\|}+\mu_{\omega\|}\sigma_{b\|}\right), \quad\mbox{for}\quad (\omega\cdot b)^{2}=1.
\end{eqnarray}
The case $\omega\cdot b=0$, is a special case of \eqref{appA7} [see \eqref{N20a}]. The corresponding energy-momentum tensor is given by \eqref{N21a}. For $\left(\omega\cdot b\right)^{2}=1$, i.e. when the magnetic field and the vorticity are either parallel or antiparallel,  $p_{6}$ is given by \eqref{appA8}.  Plugging the results for $p_{i}, i=1,\cdots,6$ into the ansatz for $\Theta^{\mu\nu}$ from \eqref{N7a}, we arrive at
\begin{eqnarray}\label{appA9}
\Theta^{\mu\nu}_{0}&=&\epsilon u^{\mu}u^{\nu}-p_{\perp}\Xi^{\mu\nu}+p_{\times}  {\omega}^{\mu}  {\omega}^{\nu}+p_{\|}b^{\mu}b^{\nu}
\nonumber\\
&&+\frac{1}{2}B^2\left(u^{\mu}u^{\nu}-\Xi^{\mu\nu}-b^{\mu}b^{\nu}+  {\omega}^{\mu}  {\omega}^{\nu}\right)\nonumber\\
&&+\left(\mu_{b\|}\sigma_{\omega\|}+\mu_{\omega\|}\sigma_{b\|}\right)b^{(\mu}  {\omega}^{\nu)},
\end{eqnarray}
where $p_{\perp}, p_{\times}$, and $p_{\|}$ are given in \eqref{N22a}.
\end{appendix}



\begin{thebibliography}{99}
\bibitem{rezzolla-book}
L.~Rezzolla andd O.~Zanotti, \textit{Relativistic Hydrodynamics}, Oxford University Press, 2013.
\bibitem{rischke-book}
G.~S.~Denicol and D.~H.~Rischke, \textit{Microscopic Foundations of Relativistic Fluid Dynamics}, Springer Nature Switzerland AG, 2021.
\bibitem{heinz2013}
U.~Heinz and R.~Snellings,
\textit{Collective flow and viscosity in relativistic heavy-ion collisions},
Ann. Rev. Nucl. Part. Sci. \textbf{63}, 123 (2013),
\href{https://arxiv.org/pdf/arXiv:1301.2826.pdf}{arXiv:1301.2826 [nucl-th]}.
\bibitem{heller2017}
W.~Florkowski, M.~P.~Heller and M.~Spalinski,
\textit{New theories of relativistic hydrodynamics in the LHC era},
Rept. Prog. Phys. \textbf{81},  046001 (2018),
\href{https://arxiv.org/pdf/arXiv:1707.02282.pdf}{arXiv:1707.02282 [hep-ph]}.
\bibitem{shen2020}
C.~Shen and L.~Yan,
\textit{Recent development of hydrodynamic modeling in heavy-ion collisions},
Nucl. Sci. Tech. \textbf{31}, 122 (2010),
\href{https://arxiv.org/pdf/arXiv:2010.12377.pdf}{arXiv:2010.12377 [nucl-th]}.
\bibitem{becattini2016}
F.~Becattini, I.~Karpenko, M.~Lisa, I.~Upsal and S.~Voloshin,
\textit{Global hyperon polarization at local thermodynamic equilibrium with vorticity, magnetic field and feed-down},
Phys. Rev. C \textbf{95}, 054902 (2017),
\href{https://arxiv.org/pdf/arXiv:1610.02506.pdf}{arXiv:1610.02506 [nucl-th]}.
\bibitem{star2017}
L.~Adamczyk \textit{et al.} [STAR],
\textit{Global $\Lambda$ hyperon polarization in nuclear collisions: evidence for the most vortical fluid},
Nature \textbf{548}, 62 (2017),
\href{https://arxiv.org/pdf/arXiv:1701.06657.pdf}{arXiv:1701.06657 [nucl-ex]}.
\bibitem{star2018}
J.~Adam \textit{et al.} [STAR],
\textit{Global polarization of $\Lambda$ hyperons in Au+Au collisions at $\sqrt{s_{_{NN}}}$ = 200 GeV},
Phys. Rev. C \textbf{98}, 014910 (2018),
\href{https://arxiv.org/pdf/arXiv:1805.04400.pdf}{arXiv:1805.04400 [nucl-ex]}.
\bibitem{star2019}
S.~Acharya \textit{et al.} [ALICE],
\textit{Evidence of spin-orbital angular momentum interactions in relativistic heavy-ion collisions},
Phys. Rev. Lett. \textbf{125}, 012301 (2020),
\href{https://arxiv.org/pdf/arXiv:1910.14408.pdf}{arXiv:1910.14408 [nucl-ex]}.
\bibitem{star2020}
J.~Adam \textit{et al.} [STAR],
\textit{Global polarization of $\Xi$ and $\Omega$ hyperons in Au+Au collisions at $\sqrt {s_{NN}}$ = 200  GeV},
Phys. Rev. Lett. \textbf{126}, 162301 (2021),
\href{https://arxiv.org/pdf/arXiv:2012.13601.pdf}{arXiv:2012.13601 [nucl-ex]}.
\bibitem{star2022}
M.~S.~Abdallah \textit{et al.} [STAR]
\textit{Pattern of global spin alignment of \ensuremath{\phi} and K$^{*0}$ mesons in heavy-ion collisions},
Nature \textbf{614}, 244 (2023),
\href{https://arxiv.org/pdf/arXiv:2204.02302.pdf}{arXiv:2204.02302 [hep-ph]}.
\bibitem{alice2022}
S.~Acharya \textit{et al.}  [ALICE],
\textit{Measurement of the J/$\psi$ polarization with respect to the event plane in Pb-Pb collisions at the LHC},
\href{https://arxiv.org/pdf/arXiv:2204.10171.pdf}{arXiv:2204.10171 [nucl-ex]}.
\bibitem{theospin2004-1}
Z.~T.~Liang and X.~N.~Wang,
\textit{Globally polarized quark-gluon plasma in non-central A+A collisions},
Phys. Rev. Lett. \textbf{94}, 102301 (2005),
[erratum: Phys. Rev. Lett. \textbf{96}, 039901 (2006)],
\href{https://arxiv.org/pdf/nucl-th/0410079.pdf}{arXiv:nucl-th/0410079 [nucl-th]}.
\bibitem{theospin2004-2}
Z.~T.~Liang and X.~N.~Wang,
\textit{Spin alignment of vector mesons in non-central A+A collisions},
Phys. Lett. B \textbf{629}, 20 (2005),
\href{https://arxiv.org/pdf/nucl-th/0411101.pdf}{arXiv:nucl-th/0411101 [nucl-th]}.
\bibitem{betz2007}
B.~Betz, M.~Gyulassy and G.~Torrieri,
\textit{Polarization probes of vorticity in heavy ion collisions},
Phys. Rev. C \textbf{76}, 044901 (2007),
\href{https://arxiv.org/pdf/arXiv:0708.0035.pdf}{arXiv:0708.0035 [nucl-th]}.
\bibitem{becattini2007}
F.~Becattini, F.~Piccinini and J.~Rizzo,
\textit{Angular momentum conservation in heavy ion collisions at very high energy},
Phys. Rev. C \textbf{77}, 024906 (2008),
\href{https://arxiv.org/pdf/arXiv:0711.1253.pdf}{arXiv:0711.1253 [nucl-th]}.
\bibitem{huang2011}
X.~G.~Huang, P.~Huovinen and X.~N.~Wang,
\textit{Quark polarization in a viscous quark-gluon plasma},
Phys. Rev. C \textbf{84}, 054910 (2011),
\href{https://arxiv.org/pdf/arXiv:1108.5649.pdf}{arXiv:1108.5649 [nucl-th]}.
\bibitem{becattini2013}
F.~Becattini, V.~Chandra, L.~Del Zanna and E.~Grossi,
\textit{Relativistic distribution function for particles with spin at local thermodynamical equilibrium},
Annals Phys. \textbf{338}, 32 (2013),
\href{https://arxiv.org/pdf/arXiv:1303.3431.pdf}{arXiv:1303.3431 [nucl-th]}.
\bibitem{rev-wang2020}
J.~H.~Gao, G.~L.~Ma, S.~Pu and Q.~Wang,
\textit{Recent developments in chiral and spin polarization effects in heavy-ion collisions},
Nucl. Sci. Tech. \textbf{31}, 90 (2020),
\href{https://arxiv.org/pdf/arXiv:2005.10432.pdf}{arXiv:2005.10432 [hep-ph]}.
\bibitem{rev-huang2020}
Y.~C.~Liu and X.~G.~Huang,
\textit{Anomalous chiral transports and spin polarization in heavy-ion collisions},
Nucl. Sci. Tech. \textbf{31}, 56 (2020),
\href{https://arxiv.org/pdf/arXiv:2003.12482.pdf}{arXiv:2003.12482 [nucl-th]}.
\bibitem{rev-lisa2020}
F.~Becattini and M.~A.~Lisa,
\textit{Polarization and vorticity in the Quark--Gluon Plasma},
Ann. Rev. Nucl. Part. Sci. \textbf{70}, 395 (2020),
\href{https://arxiv.org/pdf/arXiv:2003.03640.pdf}{arXiv:2003.03640 [nucl-ex]}.
\bibitem{becattini2022}
F.~Becattini,
\textit{Spin and polarization: a new direction in relativistic heavy ion physics},
Rept. Prog. Phys. \textbf{85}, 122301 (2022),
\href{https://arxiv.org/pdf/arXiv:2204.01144.pdf}{arXiv:2204.01144 [nucl-th]}.
\bibitem{becattini2021}
F.~Becattini, J.~Liao and M.~Lisa,
\textit{Strongly interacting matter under rotation: An introduction},
Lect. Notes Phys. \textbf{987}, 1 (2021),
\href{https://arxiv.org/pdf/arXiv:2102.00933.pdf}{arXiv:2102.00933 [nucl-th]}.
\bibitem{florkowski2017}
W.~Florkowski, B.~Friman, A.~Jaiswal and E.~Speranza,
\textit{Relativistic fluid dynamics with spin},
Phys. Rev. C \textbf{97},  041901 (2018),
\href{https://arxiv.org/pdf/arXiv:1705.00587.pdf}{arXiv:1705.00587 [nucl-th]}.
\bibitem{becattini2018}
F.~Becattini, W.~Florkowski and E.~Speranza,
\textit{Spin tensor and its role in non-equilibrium thermodynamics},
Phys. Lett. B \textbf{789}, 419-425 (2019),
\href{https://arxiv.org/pdf/arXiv:1807.10994.pdf}{arXiv:1807.10994 [hep-th]}.
\bibitem{florkowski2018}
W.~Florkowski, A.~Kumar and R.~Ryblewski,
\textit{Relativistic hydrodynamics for spin-polarized fluids},
Prog. Part. Nucl. Phys. \textbf{108}, 103709 (2019),
\href{https://arxiv.org/pdf/arXiv:1811.04409.pdf}{arXiv:1811.04409 [nucl-th]}.
\bibitem{hattori2019}
K.~Hattori, M.~Hongo, X.~G.~Huang, M.~Matsuo and H.~Taya,
\textit{Fate of spin polarization in a relativistic fluid: An entropy-current analysis},
Phys. Lett. B \textbf{795}, 100 (2019),
\href{https://arxiv.org/pdf/arXiv:1901.06615.pdf}{arXiv:1901.06615 [hep-th]}.
\bibitem{fukushima2019}
K.~Fukushima and S.~Pu,
\textit{Spin hydrodynamics and symmetric energy-momentum tensors \textendash{} A current induced by the spin vorticity \textendash{}},
Phys. Lett. B \textbf{817}, 136346 (2021),
\href{https://arxiv.org/pdf/arXiv:2010.01608.pdf}{arXiv:2010.01608 [hep-th]}.
\bibitem{florkowski2020}
S.~Bhadury, W.~Florkowski, A.~Jaiswal, A.~Kumar and R.~Ryblewski,
\textit{Relativistic dissipative spin dynamics in the relaxation time approximation},
Phys. Lett. B \textbf{814}, 136096 (2021),
\href{https://arxiv.org/pdf/arXiv:2002.03937.pdf}{arXiv:2002.03937 [hep-ph]}.
\bibitem{gale2020}
S.~Shi, C.~Gale and S.~Jeon,
\textit{From chiral kinetic theory to relativistic viscous spin hydrodynamics},
Phys. Rev. C \textbf{103},  044906 (2021),
\href{https://arxiv.org/pdf/arXiv:2008.08618.pdf}{arXiv:2008.08618 [nucl-th]}.
\bibitem{stephanov2020}
S.~Li, M.~A.~Stephanov and H.~U.~Yee,
\textit{Nondissipative second-order transport, spin, and pseudogauge transformations in hydrodynamics},
Phys. Rev. Lett. \textbf{127}, 082302 (2021),
\href{https://arxiv.org/pdf/arXiv:2011.12318.pdf}{arXiv:2011.12318 [hep-th]}.
\bibitem{yarom2021}
A.~D.~Gallegos, U.~G\"ursoy and A.~Yarom,
\textit{Hydrodynamics of spin currents},
SciPost Phys. \textbf{11}, 041 (2021),
\href{https://arxiv.org/pdf/arXiv:2101.04759.pdf}{arXiv:2101.04759 [hep-th]}.
\bibitem{wang2021}
H.~H.~Peng, J.~J.~Zhang, X.~L.~Sheng and Q.~Wang,
\textit{Ideal spin hydrodynamics from the Wigner function approach},
Chin. Phys. Lett. \textbf{38}, 116701 (2021),
\href{https://arxiv.org/pdf/arXiv:2107.00448.pdf}{arXiv:2107.00448 [hep-th]}.
\bibitem{nora2022}
N.~Weickgenannt, D.~Wagner, E.~Speranza and D.~H.~Rischke,
\textit{Relativistic second-order dissipative spin hydrodynamics from the method of moments},
Phys. Rev. D \textbf{106}, 096014 (2022),
\href{https://arxiv.org/pdf/arXiv:2203.04766.pdf}{arXiv:2203.04766 [nucl-th]}.
\bibitem{mishra2022}
G.~Sarwar, M.~Hasanujjaman, J.~R.~Bhatt, H.~Mishra and J.~e.~Alam,
\textit{Causality and stability of relativistic spin hydrodynamics,}
Phys. Rev. D \textbf{107}, 054031 (2023),
\href{https://arxiv.org/pdf/arXiv:2209.08652.pdf}{arXiv:2209.08652 [nucl-th]}.
\bibitem{ryblewski2022}
A.~Daher, A.~Das and R.~Ryblewski,
\textit{Stability studies of first-order spin-hydrodynamic frameworks},
Phys. Rev. D \textbf{107},  054043 (2023),
\href{https://arxiv.org/pdf/arXiv:2209.10460.pdf}{arXiv:2209.10460 [nucl-th]}.
\bibitem{pu2023}
X.~Q.~Xie, D.~L.~Wang, C.~Yang and S.~Pu,
\textit{Causality and stability analysis for the minimal causal spin hydrodynamics},
\href{https://arxiv.org/pdf/arXiv:2306.13880.pdf}{arXiv:2306.13880 [hep-ph]}.
\bibitem{nora2023}
N.~Weickgenannt,
\textit{Linearly stable and causal relativistic first-order spin hydrodynamics},
\href{https://arxiv.org/pdf/arXiv:2307.13561.pdf}{arXiv:2307.13561 [nucl-th]}.
\bibitem{hu2022}
J.~Hu,
\textit{Cross effects in spin hydrodynamics: Entropy analysis and statistical operator},
Phys. Rev. C \textbf{107},  024915 (2023),
\href{https://arxiv.org/pdf/arXiv:2209.10979.pdf}{arXiv:2209.10979 [nucl-th]}.
\bibitem{becattini2023}
F.~Becattini, A.~Daher and X.~L.~Sheng,
\textit{Entropy current and entropy production in relativistic spin hydrodynamics},
\href{https://arxiv.org/pdf/arXiv:2309.05789.pdf}{arXiv:2309.05789 [nucl-th]}.
\bibitem{huang2022}
Z.~Cao, K.~Hattori, M.~Hongo, X.~G.~Huang and H.~Taya,
\textit{Gyrohydrodynamics: Relativistic spinful fluid with strong vorticity},
PTEP \textbf{2022}, 071D01 (2022),
\href{https://arxiv.org/pdf/2205.08051.pdf}{arXiv:2205.08051 [hep-th]}.
\bibitem{warringa2007}
D.~E.~Kharzeev, L.~D.~McLerran and H.~J.~Warringa,
\textit{The Effects of topological charge change in heavy ion collisions: 'Event by event P and CP violation'},
Nucl. Phys. A \textbf{803}, 227 (2008),
\href{https://arxiv.org/pdf/0711.0950.pdf}{arXiv:0711.0950 [hep-ph]}.
\bibitem{skokov2009}
V.~Skokov, A.~Y.~Illarionov and V.~Toneev,
\textit{Estimate of the magnetic field strength in heavy-ion collisions},
Int. J. Mod. Phys. A \textbf{24}, 5925 (2009),
\href{https://arxiv.org/pdf/0907.1396.pdf}{arXiv:0907.1396 [nucl-th]}.
\bibitem{rev-huang2015}
X.~G.~Huang,
\textit{Electromagnetic fields and anomalous transports in heavy-ion collisions --- A pedagogical review},
Rept. Prog. Phys. \textbf{79}, 076302 (2016),
\href{https://arxiv.org/pdf/1509.04073.pdf}{arXiv:1509.04073 [nucl-th]}.
\bibitem{sadooghi2010}
S.~Fayazbakhsh and N.~Sadooghi,
\textit{Phase diagram of hot magnetized two-flavor color superconducting quark matter,}
Phys. Rev. D \textbf{83}, 025026 (2011),
\href{https://arxiv.org/pdf/1009.6125.pdf}{arXiv:1009.6125 [hep-ph]}.
\bibitem{sadooghi2012}
S.~Fayazbakhsh, S.~Sadeghian and N.~Sadooghi,
\textit{Properties of neutral mesons in a hot and magnetized quark matter},
Phys. Rev. D \textbf{86}, 085042 (2012),
\href{https://arxiv.org/pdf/1206.6051.pdf}{arXiv:1206.6051 [hep-ph]}.
\bibitem{sadooghi2013}
S.~Fayazbakhsh and N.~Sadooghi,
\textit{Weak decay constant of neutral pions in a hot and magnetized quark matter},
Phys. Rev. D \textbf{88}, 065030 (2013),
\href{https://arxiv.org/pdf/1306.2098.pdf}{arXiv:1306.2098 [hep-ph]}.
\bibitem{sadooghi2014}
S.~Fayazbakhsh and N.~Sadooghi,
\textit{Anomalous magnetic moment of hot quarks, inverse magnetic catalysis, and reentrance of the chiral symmetry broken phase},
Phys. Rev. D \textbf{90}, 105030 (2014),
\href{https://arxiv.org/pdf/1408.5457.pdf}{arXiv:1408.5457 [hep-ph]}.
\bibitem{sadooghi2015}
N.~Sadooghi and F.~Taghinavaz,
\textit{Magnetized plasminos in cold and hot QED plasmas},
Phys. Rev. D \textbf{92}, 025006 (2015),
\href{https://arxiv.org/pdf/1504.04268.pdf}{arXiv:1504.04268 [hep-ph]}.
\bibitem{taghinavaz2016}
N.~Sadooghi and F.~Taghinavaz,
\textit{Dilepton production rate in a hot and magnetized quark-gluon plasma},
Annals Phys. \textbf{376}, 218 (2017),
\href{https://arxiv.org/pdf/1601.04887.pdf}{arXiv:1601.04887 [hep-ph]}.
\bibitem{sadooghi2020}
S.~M.~A.~Tabatabaee and N.~Sadooghi,
\textit{Wigner function formalism and the evolution of thermodynamic quantities in an expanding magnetized plasma},
Phys. Rev. D \textbf{101}, 076022 (2020),
\href{https://arxiv.org/pdf/2003.01686.pdf}{arXiv:2003.01686 [hep-ph]}.
\bibitem{sadooghi2019}
N.~Sadooghi and S.~M.~A.~Tabatabaee,
\textit{Paramagnetic squeezing of a uniformly expanding quark-gluon plasma in and out of equilibrium},
Phys. Rev. D \textbf{99},  056021 (2019),
\href{https://arxiv.org/pdf/1901.06928.pdf}{arXiv:1901.06928 [nucl-th]}.
\bibitem{rischke2009}
X.~G.~Huang, M.~Huang, D.~H.~Rischke and A.~Sedrakian,
\textit{Anisotropic hydrodynamics, bulk viscosities and R-modes of strange quark stars with strong magnetic fields},
Phys. Rev. D \textbf{81}, 045015 (2010),
\href{https://arxiv.org/pdf/0910.3633.pdf}{arXiv:0910.3633 [astro-ph.HE]}.
\bibitem{rischke2011}
X.~G.~Huang, A.~Sedrakian and D.~H.~Rischke,
\textit{Kubo formulae for relativistic fluids in strong magnetic fields},
Annals Phys. \textbf{326}, 3075 (2011),
\href{https://arxiv.org/pdf/1108.0602.pdf}{arXiv:1108.0602 [astro-ph.HE]}.
\bibitem{sadooghi2016}
N.~Sadooghi and S.~M.~A.~Tabatabaee,
\textit{The effect of magnetization and electric polarization on the anomalous transport coefficients of a chiral fluid},
New J. Phys. \textbf{19}, 053014 (2017),
\href{https://arxiv.org/pdf/1612.02212.pdf}{arXiv:1612.02212 [hep-th]}.
\bibitem{iqbal2016}
S.~Grozdanov, D.~M.~Hofman and N.~Iqbal,
\textit{Generalized global symmetries and dissipative magnetohydrodynamics},
Phys. Rev. D \textbf{95},  096003 (2017),
\href{https://arxiv.org/pdf/arXiv:1610.07392.pdf}{arXiv:1610.07392 [hep-th]}.
\bibitem{hongo2020}
M.~Hongo and K.~Hattori,
\textit{Revisiting relativistic magnetohydrodynamics from quantum electrodynamics},
JHEP \textbf{02}, 011 (2021),
\href{https://arxiv.org/pdf/arXiv:2005.10239.pdf}{arXiv:2005.10239 [hep-th]}.
\bibitem{hongo2022}
K.~Hattori, M.~Hongo and X.~G.~Huang,
\textit{New developments in relativistic magnetohydrodynamics},
Symmetry \textbf{14}, 1851 (2022),
\href{https://arxiv.org/pdf/2207.12794.pdf}{arXiv:2207.12794 [hep-th]}.
\bibitem{florkowski2022}
S.~Bhadury, W.~Florkowski, A.~Jaiswal, A.~Kumar and R.~Ryblewski,
\textit{Relativistic spin magnetohydrodynamics},
Phys. Rev. Lett. \textbf{129}, 192301 (2022),
\href{https://arxiv.org/pdf/2204.01357.pdf}{arXiv:2204.01357 [nucl-th]}.
\bibitem{shokri2022}
R.~Singh, M.~Shokri and S.~M.~A.~Tabatabaee~Mehr,
\textit{Relativistic hydrodynamics with spin in the presence of electromagnetic fields},
Nucl. Phys. A \textbf{1035}, 122656 (2023),
\href{https://arxiv.org/pdf/arXiv:2202.11504.pdf}{arXiv:2202.11504 [hep-ph]}.
\bibitem{buzzegoli2022}
M.~Buzzegoli,
\textit{Spin polarization induced by magnetic field and the relativistic Barnett effect},
Nucl. Phys. A \textbf{1036}, 122674 (2023),
\href{https://arxiv.org/pdf/arXiv:2211.04549.pdf}{arXiv:2211.04549 [nucl-th]}.
\bibitem{pu2022}
H.~H.~Peng, S.~Wu, R.~J.~Wang, D.~She and S.~Pu,
\textit{Anomalous magnetohydrodynamics with temperature-dependent electric conductivity and application to the global polarization},
Phys. Rev. D \textbf{107}, 096010 (2023),
\href{https://arxiv.org/pdf/arXiv:2211.11286.pdf}{arXiv:2211.11286 [hep-ph]}.
\bibitem{glibert-strang}
G. Strang, \textit{Introduction to Linear Algebra}, Sixth edition, Wellesley-Cambridge Press, 2023.
\bibitem{shokri2021}
M.~Kiamari, M.~Rahbardar, M.~Shokri and N.~Sadooghi,
\textit{Anomalous Hall instability in the Chern-Simons magnetohydrodynamics},
Phys. Rev. D \textbf{104}, 076023 (2021),
\href{https://arxiv.org/pdf/2102.11695.pdf}{arXiv:2102.11695 [hep-th]}.
\bibitem{shokri2023}
M.~Shokri and D.~H.~Rischke,
\textit{Linear stability analysis in inhomogeneous equilibrium configurations},
\href{https://arxiv.org/pdf/2309.07003.pdf}{arXiv:2309.07003 [physics.flu-dyn]]}.







\end{thebibliography}
\end{document}